\documentclass[prb,twocolumn,aps,showpacs,amssymb,amsfonts,amsmath]{revtex4}
\usepackage[dvips]{graphicx}
\usepackage{color}
\usepackage{epsfig}
\usepackage[normalem]{ulem}	% to enable cross-out of sentences or words
\newcommand{\bom}{{\mbox{\boldmath $\omega$}}}

\definecolor{g-blue}{rgb}{0.83,0.95,1}
\definecolor{g-yellow}{rgb}{1,1,0.7}
\definecolor{g-green}{rgb}{0.9,1,0.9}
\definecolor{green}{rgb}{0,0.6,0}
\definecolor{cyan}{rgb}{0,0.7,0.7}
\definecolor{black}{rgb}{0,0,0}
\definecolor{grey}{rgb}{0.4 ,0.4 ,0.4 }

\def\blue#1{\textcolor{blue}{#1}}

\newcommand{\B}[1]{{\bf{#1}}}%% Bold Roman & Greek Lower & Upper Case
\renewcommand{\sb}[1]{_{\text {#1}}}  %% sub-   for lower case
\renewcommand{\sp}[1]{^{\text {#1}}}  %% super- for lower case
\newcommand{\Sp}[1]{^{^{\text {#1}}}} %% Super- for Upper case
\def\Sb#1{_{\scriptscriptstyle\rm{#1}}}

\def\Fbox#1{\vskip1ex\hbox to 8.5cm{\hfil\fboxsep0.3cm\fbox{%
  \parbox{8.0cm}{#1}}\hfil}\vskip1ex\noindent}  %%  {TEXT} in BOX
\begin{document}

\title{\blue{Turbulent velocity spectra in a quantum fluid: \\
experiments,  numerics and models}}

\author{Carlo F. Barenghi$^\dag$,   Victor L'vov$^*$, and Philippe-E. Roche$^\ddag$}

\address{ $^\dag$Joint Quantum Centre Durham-Newcastle and School
of Mathematics and Statistics, Newcastle University, Newcastle upon Tyne,
NE1 7RU, United Kingdom, \\
  $^*$Weizmann Institute of Science, Rehovot 76100, Israel,\\
$^\ddag$Institut NEEL, CNRS/UJF, F-38042 Grenoble 9,
France}

\begin{abstract}
Turbulence in superfluid helium is unusual and
presents a challenge to fluid dynamicists because it consists of
two coupled, inter penetrating turbulent fluids: the first is inviscid with
quantised vorticity, the second is viscous with continuous vorticity.
Despite this double nature, the observed spectra of the superfluid
turbulent velocity
at sufficiently large length scales are
similar to those o ordinary turbulence.
We present experimental, numerical and theoretical results which
explain these similarities, and illustrate the limits
of our present understanding of superfluid turbulence at smaller scales.
 \end{abstract}
\maketitle

%%%%%%%%%%%%%%%%%%%%%%%%%%%%%%%%%%%%%%%%%%%%%%%%%%%%%%%%%%%%%%%%%

\section{ Motivations}

If cooled below a critical temperature
($T_{ \lambda}   \approx 2.18 \,$K in $^4$He and
$T_{\rm c}\approx  10^{-3}\,K$ in at $^3$He
\footnote{Hereafter by $^3$He we mean the B-phase of $^3$He}
at saturated
vapour pressure), liquid helium undergoes Bose-Einstein condensation
\cite{Annett}, becoming a quantum fluid and
demonstrating superfluidity (pure inviscid flow).
Besides the lack of viscosity, another major difference between
superfluid helium and ordinary (classical) fluids such as water or air is
that, in helium, vorticity is constrained to vortex line singularities of fixed
circulation $\kappa=h/M$,  where $h$ is Planck's
constant,  and  $M$ is the mass of the relevant boson
(in the most common isotope $^4$He, $M=m_4$, the mass of an atom; in the
rare isotope $^3$He, $M=2\,m_3$, the mass of
a Cooper pair). These vortex lines are
essentially one-dimensional space curves, like the vortex lines of
fluid dynamics textbooks; for example, in $^4$He the vortex core radius
$\xi \approx 10^{-10}$m is comparable to the inter atomic distance.
This quantisation of the circulation thus results in the appearance  of
another characteristic length scale:
the mean separation between vortex lines, $\ell$.
In typical experiments (both in $^4$He and $^3$He) $\ell$ is orders
of magnitude
smaller than the outer scale of turbulence $D$ (the scale of the
largest eddies) but is also orders of magnitudes larger than $\xi$.

There is a growing consensus \cite{Skrbek-Sreeni-2012}
that on length scales much larger than $\ell$ the properties of
superfluid turbulence
are similar to those of classical turbulence if excited similarly,
for example by a moving grid. The idea is that motions at scales
$R \gg \ell$ should  involve at least
a partial polarization \cite{Hulton,Laurie-2012}
of vortex lines and their organisation into vortex bundles
which, at such large scales, should mimic continuous hydrodynamic eddies.
Therefore one expects a classical Richardson-Kolmogorov energy cascade,
with larger ``eddies'' breaking into smaller ones.
The spectral signature of this classical cascade
is indeed observed experimentally in superfluid helium.
%see Figs.~\ref{f:2}-A and B below.
In the absence of viscosity, in superfluid
turbulence the kinetic energy should cascade downscale without loss, until it
reaches the small scales where the quantum discreteness of vorticity
is important. It is also believed that at this point the Richardson-Kolmogorov
eddy-dominated cascade should be replaced by a second cascade which arises from
the nonlinear interaction of Kelvin waves (helical perturbation of the
vortex lines) on individual vortex lines.
This Kelvin wave cascade should take the energy further
downscale where it is radiated away by thermal
quasi particles (phonons and  rotons  in $^4$He).

Although this scenario seems quite reasonable, crucial
details are yet to be established. Our understanding of superfluid
turbulence at scales of the order of $\ell$ is still at infancy stage,
and what happens at scales below $\ell$ is a question of intensive debates.
The ``quasi-classical" region of scales, $R\gg \ell$, is
better understood, but still less than classical hydrodynamic turbulence.
The main reason is that at nonzero temperatures (but still below the critical
temperature $T_{ \lambda}$), superfluid helium is a two-fluid system. According to the
theory of Landau and Tisza~\cite{Donnelly}, it consists of two
inter--penetrating components: the inviscid superfluid, of density $\rho\sb s$
and velocity $\B u \sb s$ (associated to the quantum ground state), and the
viscous normal fluid, of density $\rho_n$ and velocity $\bm u \sb n$
(associated to thermal excitations). The normal fluid carries the entropy
$S$ and the viscosity $\mu$ of the entire liquid. In the presence of superfluid
vortices these two components interact
via a mutual friction force\cite{BDV1982}.  The  total helium density
$\rho=\rho\sb s+\rho_n \approx~ 145~\rm kg/m^3$ is practically temperature
independent, while the superfluid fraction $\rho\sb s/\rho$ is zero
at $T=T_\lambda$, but rapidly increases if $T$ is lowered
(it becomes 50\% at $T \approx 2~\rm K$,   83\%
at $T \approx 1.6~\rm K$  and 95\% at $T \approx 1.3~\rm K$ \cite{DB}).
The normal fluid is essentially negligible below $1~\rm K$.
One would therefore expect classical behaviour only in the high temperature
limit $T \to T_{\lambda}$, where the normal fluid must
energetically dominate the dynamics. Experiments show that this is not
the case, thus raising the interesting problem
of ``double-fluid" turbulence which we study here.

The aim of this article is to present the current
state of the art in this intriguing problem, clarify common features of
turbulence in classical and quantum fluids, and highlight their differences.
To achieve our aim we shall overview and combine experimental,
theoretical and numerical results in the simplest possible (and, probably,
the most fundamental) case of homogeneous, isotropic turbulence,
away from boundaries and maintained in a statistical steady state by
continuous mechanical forcing.
The natural tools to study homogeneous isotropic
turbulence are spectral, thus we shall consider
the velocity spectrum (also known as the energy spectrum)
and attempt to give a physical explanation for the observed phenomena.

%%%%%%%%%%%%%%%%%%%%%%%%%%%%%%%%%%%%%%%%%%%%%%%%%%%%%%%%%%%%%%%%%

\section{\label{s:background}
%2. \magenta{Physical background of   superfluid turbulence in $^4$He}}
Classical  \emph{vs.}  superfluid turbulence}

We recall that
ordinary incompressible viscous flows are described by the
Navier-Stokes equation
\begin{equation}\label{NavierStokes}
\Big[\frac{\partial \,\bf u}{\partial t}+ (\B u \cdot \B
\nabla)\B u \Big] =
- \frac{1}{\rho}{\B \nabla p}+\nu {\B \nabla^2} \B u,
\end{equation}
and the solenoidal condition $\B \nabla \cdot \B u=0$
for the velocity field $\B u$,
where $p$ is the pressure, $\rho$ the density,
and $\nu=\mu/\rho$ the kinematic viscosity.
The dimensionless parameter that determines the properties
of hydrodynamic turbulence is the Reynolds number Re$=V D/\nu$.
The Reynolds number estimates  the ratio of nonlinear and viscous terms in
Eq.~\eqref{NavierStokes} at the outer length scale $D$
(typically the size  of a streamlined body), where
$V$ is  the root mean square
turbulent velocity fluctuation.
In fully developed
turbulence (Re$\gg 1$) the $D$-scale eddies are unstable and
give birth to smaller scale eddies,  which, being unstable,
generate further smaller eddies, and so on.
This process is the
Richardson-Kolmogorov energy cascade toward eddies of scale $\eta$,
defined as the length scale at which
the nonlinear and viscous forces in Eq.~\eqref{NavierStokes}
approximately balance each other.
$\eta$-scale eddies are
stable and their energy is dissipated into heat by viscous forces.
The hallmark feature of fully developed turbulence is
 the coexistence of
eddies of all scales from $D$ to $\eta \simeq D {\rm Re}^{-3/4}\ll D$ with
universal statistics; the range of length scales
$D\ll R \ll \eta$ where both energy pumping and
dissipation can be ignored
is called the inertial range.

In the study of homogeneous turbulence it is customary
to consider the energy density per unit mass $E(t)$
(of dimensions $\rm m^2/s^2$).
In the isotropic case the energy distribution between eddies of scale $R$
can be characterized by the one--dimensional energy
spectrum $E(k,t)$ of dimensions $\rm m^2/s^2$) with wavenumber defined as
$k=2\pi/R$ (or as $k=1/R$), normalized such that\\
% \begin{subequations}\label{En}\begin{equation}\label{EnA}
$$   E(t)=\frac{1}{\cal V} \int \frac{1}{2} {\B u}^2 d{\cal V}=
\int_0^\infty E(k,t)d k,
$$
where ${\cal V}$ is volume.
In  the inviscid limit, $E(t)$ is a conserved quantity ($d E(t)/ dt =0$), thus
$E(k,t)$ satisfies the continuity equation
\begin{equation}\label{EnB}
\frac{\partial E(k,t)}{\partial t}
+\frac{\partial \varepsilon (k,t)}{\partial k}=0,
\end{equation}

\noindent
where $\varepsilon(k,t)$ is the energy flux in spectral space (of dimensions $\rm m^2/s^3$).
In the stationary case,  energy spectrum and energy flux are
$t$--independent, thus
Eq.~\eqref{EnB} immediately dictates that the energy flux
$\varepsilon$ is $k$-independent.
Assuming that this constant
$\varepsilon$ is the only relevant characteristics of
turbulence in the inertial interval and using dimensional reasoning,
in 1941 Kolmogorov and (later) Obukhov suggested that the energy spectrum is
\begin{equation}\label{K41}
E\sb{K41} (\varepsilon,k)=C\Sb{K41} \varepsilon  ^{2/3} k^{-5/3},
\end{equation}
where the (dimensionless) Kolmogorov constant is approximately
$C\Sb{K41} \approx 1 $.
This is the celebrated Kolmogorov-Obukhov $5/3$ law (KO--41),
verified in experiments and numerical simulations of Eq.~\eqref{NavierStokes};
it states in particular that in incompressible,
steady, homogeneous, isotropic turbulence, the distribution of
kinetic energy over the wavenumbers is $E(k) \propto k^{-5/3}$.

In the inviscid  limit the energy flux goes to smaller and
smaller scales, reaching finally the interatomic scale and accumulating
there.  To describe this effect,  Leith~\cite{Leith67} suggested to replace
the algebraic relation~\eqref{K41} between $\varepsilon(k)$ and $E(k)$  by
the differential form:
\begin{equation}%%
  \label{Leith-n}%%
  \varepsilon (k) = -{1\over 8} \, \sqrt{k^{11}  E (k)}\  {d\,  \over d k}
\left( \frac{ E (k) }{  k^2} \right) \ .
\end{equation}
This approximation dimensionally coincides with Eq.~\eqref{K41},
 but the derivative  $d[ E (k)/k^2]/dk$
guarantees that $\varepsilon (k)=0$ if $ E (k) \propto k^2$.
The numerical factor $1/8$, suggested in \cite{Nazar-Leith},
fits the experimentally observed value
%of the Kolmogorov constant
$C\Sb {K41}=(24/11)^{2/3}\approx 1.7 $ in Eq.~\eqref{K41}.

\begin{figure*}
\begin{tabular}{|c|c|c|}
  \hline
  A & B & C \\
 \includegraphics[width=.29 \textwidth]{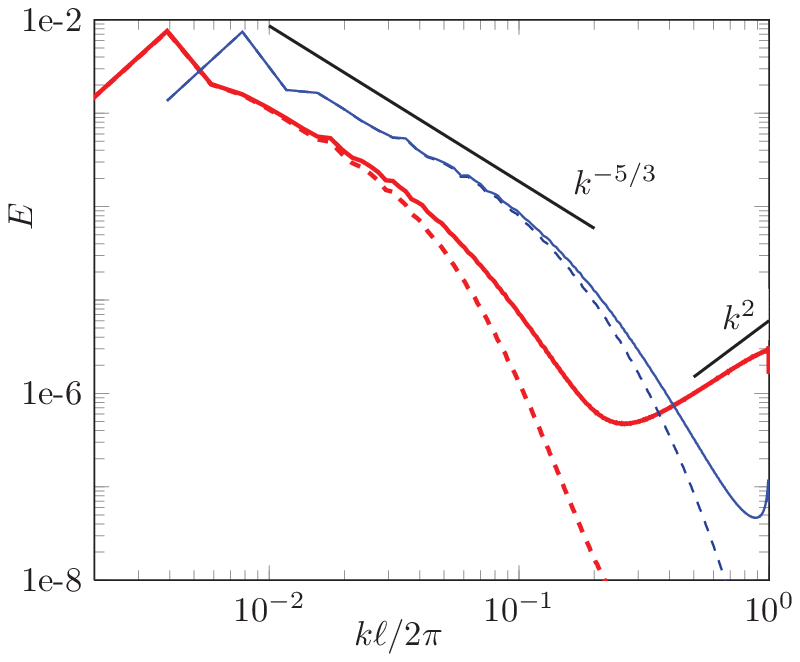} & \includegraphics[width=.32 \textwidth]{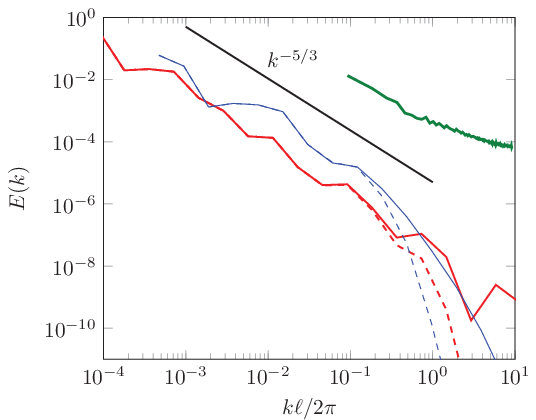} & \includegraphics[width=.34 \textwidth]{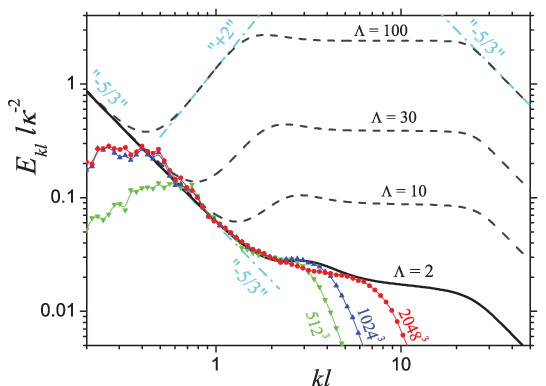} \\
  \hline
\end{tabular}
\caption{\label{f:3} Color online.
Numerical superfluid energy spectra vs normalised wavenumber for three
hierarchical levels
of motion Eqs.~\eqref{NSE}, \eqref{VFM} and \eqref{GPE}:
Panel A: Superfluid (solid line) and normal fluid (dashed line)
energy spectra from simulation of the HVBK
equations~\eqref{NSE} at $T=1.15~\rm K$ (red) and $\simeq$2.16 K (blue)
with truncation of phase space beyond the intervortex
scale \cite{Salort-EPL-2012}.
Panel B:  Superfluid energy spectrum from VFM simulations ~\eqref{VFM}
at $T=2.164~\rm K$
(solid green line) \cite{Sherwin-2013}
with synthetic turbulence prescribed for the normal
component.
Superfluid (red/blue solid line) and normal fluid (red/blue dashed line)
energy spectra from shell model simulation of the HVBK equations
at 1.44 K (red) and 2.157 K (blue) \cite{Wacks-2011}.
Panel C: Superfluid energy spectrum from GPE simulations~\eqref{GPE}
complemented by dissipation at high $k$ \cite{Sasa-2011}.
The numerical resolution is 2048$^3$ (red line), 1024$^3$ (dashed blue line)
 and 512$^3$ (green dots). The intervortex distance $\ell$ results from a
fit of the data (see original publication). In all panels,
the normalisation of the x-axis (wavevector $k$) highlights
the mean intervortex distance $\ell$.
Black dashed lines show analytical predictions of the bottleneck
\cite{LNR-2} discussed in Sec.~8A with different
$\Lambda=\ln (\ell/\xi)$. The black solid line with $\Lambda=2$
corresponds to the simulation in Ref.~\cite{Sasa-2011}.
The dashed cyan lines show (from the left) the KO--41  (-5/3) scaling,
the energy equipartition scaling ($+2$)  and, at the right,
the (-5/3) LN spectrum~Eq.~\eqref{LN}.}
\end{figure*}

A generic energy spectrum with a constant energy flux was found in
\cite{Nazar-Leith} as a solution to the
equation ~$\varepsilon(k) = \varepsilon$  constant: %%
\begin{eqnarray}%%
  \label{flux1}%%
  E (\varepsilon,k) = C\Sb{K41} \frac{\varepsilon ^{2/3}} {k^{5/3}}T\sb{eq}(k)\,, \\ \nonumber T\sb{eq}(k)=\left[ 1+\Big ( \frac{k}{k\sb{eq}}\Big)^{11/2}  \right]^{2/3}.%%
\end{eqnarray} %%
Notice that at low $k$, Eq.~\eqref{flux1} coincides with
KO--41, while for $k \gg k\sb{eq}$
it describes a thermalized  part of the spectrum, $E(k)\propto k^2$,
with equipartition of energy (shown by the solid
black line at the right of in Fig.~\ref{f:3}A, and, underneath in the same
figure, by the solid red line,
although the latter occurs in slightly different contexts)
\footnote{
In the simulations shown in Fig.~\ref{f:3}A, the energy flux
$\epsilon(k)$ is not preserved along the cascade,
but continuously decreases due to dissipation and ultimately vanishes
at the maximum $k$.}.

We shall have also to keep in mind that although
Eq.~\eqref{K41} is  an important result of classical
turbulence theory, it presents only the very beginning of the story. In particular,  its well known \cite{Frisch} that in the inertial range, the turbulent velocity field is
not self--similar, but shows intermittency effects which modify the KO--41
scenario.

In this paper we apply these ideas
to superfluid helium, explain how to overcome technical
difficulties to  measure the energy spectrum near
absolute zero, and
draw the attention to three conceptual differences
between classical hydrodynamic turbulence and turbulence in
superfluid $^4$He.

The first difference is that the quantity which (historically)
is most easily and most frequently detected in turbulent liquid helium
is not the superfluid velocity but rather the vortex line density
$\cal L$, defined as the superfluid
vortex length per unit volume; in most experiments (and numerical
simulations) this volume is the entire cell (or computational box)
which contains the helium.
This scalar quantity $\cal L$ has no analogy in classical fluid mechanics and
should not be confused with the vorticity, whose spectrum, in the
classical KO--41 scenario, scales as $k^{1/3}$ correspondingly to
the $k^{-5/3}$ scaling of the velocity. Notice that
in a superfluid the vorticity is zero everywhere except
on quantized vortex lines.  In  order
to use as much as possible the toolkit of ideas and
methods of classical hydrodynamics, we shall
define in the next sections
an "effective" superfluid vorticity field $\bom_s$; this definition
(which indeed \cite{Baggaley-structures}
yields the classical $k^{1/3}$ vorticity
scaling corresponding to the $k^{-5/3}$ velocity scaling)
is possible on scales that exceed the mean intervortex scale $\ell$,
provided  that
the vortex lines contained in a fluid parcel are sufficiently polarized.
This procedure
opens the way for a possible identification of "local" values of
${\cal L}(\B r,t)$ with the magnitude $\vert \bom_s \vert$
of the vector field $\bom_s$.

The second difference is that liquid helium is a two fluid
system, and we expect both superfluid and normal fluid to be
turbulent. This makes the problem of superfluid turbulence
much richer than classical turbulence, but the analysis becomes more
involved. For example, the existence of the intermediate scale $\ell$
makes it impossible to apply arguments of scale invariance to the entire
inertial interval and calls for its separation into three ranges.
The first is a ``hydrodynamic'' region of scales $\ell \ll R \ll D$
(corresponding to $k_D \ll k \ll k_{\ell}$ in k-space where
$k_D=2 \pi/D$ and $k_{\ell}=2 \pi/\ell$),
which is similar (but not equal) to the classical inertial range;
the second is a ``Kelvin wave region''
$\xi \ll R \ll \ell$
where energy is transferred further to smaller scales
\footnote{
Actually phonon emission will terminate the Kelvin
cascade at scales $R \sim 100 \xi$ in $^4$He\cite{Vinen2001}.
}
by interacting Kelvin waves (helix-like deformations of
the vortex lines). In the third,  intermediate region $R\approx \ell$,
the energy flux is caused probably by vortex reconnections.

Finally, the third difference is that mutual friction between normal and
superfluid components leads to (dissipative) energy exchange between them in
either direction.

Studies of classical turbulence are solidly based on the
Navier-Stokes Eq.~\eqref{NavierStokes}. Unfortunately, there are no
well established equations of motion for $^4$He in the presence
of superfluid vortices. We have only models at
different levels of description (for an overview see Sec.~\ref{s:theory}).
All these issues make the problem of superfluid turbulence very interesting
from a fundamental view point, simultaneously creating serious problems
in  experimental, numerical and analytical studies.

%\Fbox{OUR ORIGINAL SUGGESTIONS WERE :
%\blue{occurs in a different context discussed below.}
%\red{disagree-VL, suggest to replace by:} \green{as appears
%in the solid red line spectrum of Fig.~\ref{f:3}A.
%Physical reason for this thermalization will be discussed below.}
%}}}.

%%%%%%%%%%%%%%%%%%%%%%%%%%%%%%%%%%%%%%%%%%%%%%%%%%%%%%%%%%%
\begin{figure*}
\begin{tabular}{|c|c|}
  \hline
 A & B \\
  \includegraphics[width=.73\textwidth]{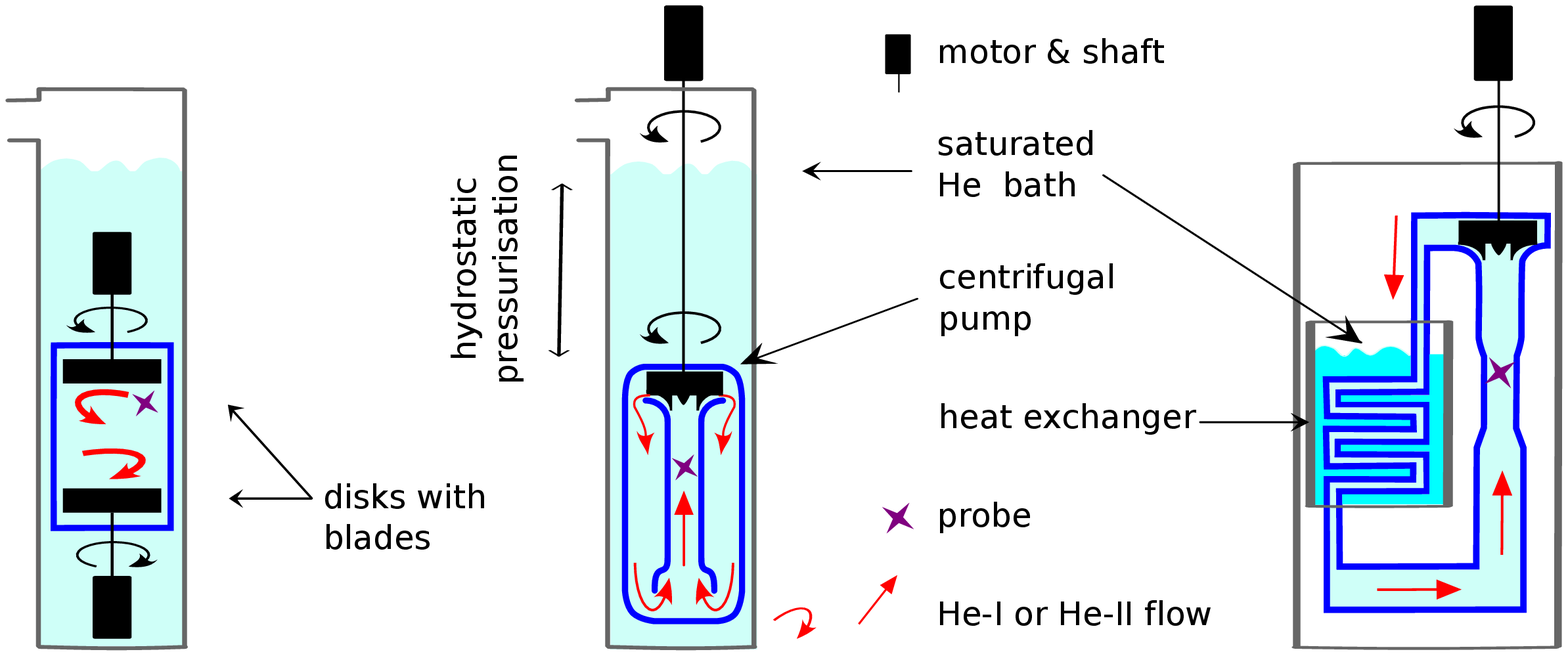} &
   \includegraphics[width=.25\textwidth]{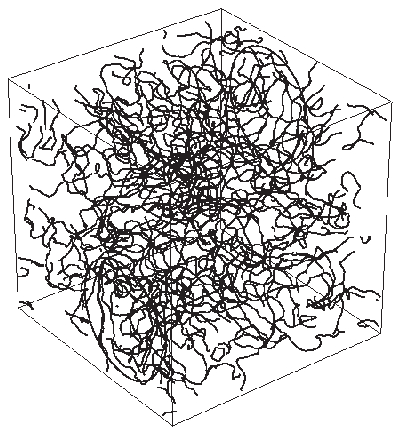} \\
  \hline
\end{tabular}
\caption{\label{f:1}
 Panel A: Examples of flows sustaining steady
turbulence in superfluid $^4$He for which spectral
measurements were performed.
From left to right : Von Karman flows (\cite{Tabeling,Schmoranzer-2009,SHREK}),
wind-tunnels (\cite{Salort-EPL-2012,Roche-2007}) and pressurised
circulator cooled through a heat exchanger \cite{Salort-2010}.
Panel B:  Snapshot of a vortex tangle calculated
using the Vortex Filament Method (VFM)
in a periodic box \cite{Sherwin-2013}.
}
\end{figure*}

%%%%%%%%%%%%%%%%%%%%%%%%%%%%%%%%%%%%%%%%%%%%%%%%%%%%%%%%%%%%%%%%%

\section{\label{s:exp} Experiments: flows, probes and spectra}

In this section we shall limit our discussion to experimental techniques
for $^4$He. The methods used in $^3$He, at temperatures which are
one thousand times smaller, are rather different \cite{Fisher-Udine},
and we shall only
cite the results in $^3$He which are directly relevant to our aim.

Possibly the simplest method to generate turbulence in $^4$He
is the application of a temperature gradient which creates a flow
of the normal component carrying heat from the hot to the cold plate;
this flow is compensated by the counterflow of the
superfluid component in the opposite direction which maintains a zero mass flux.
This form of heat conduction, called thermal counterflow,
is unlike what happens in ordinary fluids. Moreover,
under thermal drive, the energy pumping is dominated by the intervortex
length scale $\ell$ and  according to numerical simulations
there is no inertial interval in which the energy flux scales
over the wavenumbers as in the KO--41 scenario \cite{Sherwin-2012}.
The resulting
``quantum" superfluid turbulence \cite{Nemirovskii:PhysRep2013}
is thus very different from classical turbulence at large level of drive
 and will not be discussed here.

From the experimental point of view, the generation of turbulence by mechanical means
(more similar to what is done in the study of ordinary turbulence) is not as
straightforward. Nevertheless, the literature reports a number of successful approaches,
which can be classified into three main categories: (i) flows driven by vibrating objects,
(ii) one-shot-flows driven by single-stroke-bellows, towed grids or spin-up/down of
the container, and (iii) flows continuously driven by propellers.
Most efforts in characterising the turbulent fluctuations
have focused on the third category. The reason is simple: the resulting turbulent flows
do not suffer from the lack of homogeneity and isotropy which is typical of
the flows generated by vibrating objects, and allow better statistical convergence
(and improved stationarity) than measurements in non-stationary flows.

To illustrate the different cryogenics experimental set-ups,
it is useful to distinguish between the two liquid phases of $^4$He:
liquid helium~I (He-I) and helium~II (He-II), respectively
above $T_{\lambda}$ and below $T_{\lambda}$. The former is an ordinary
viscous fluid, the latter is the quantum fluid of interest here.
Since He-II is created by cooling He-I, in most cases the same
apparatus or experimental technique can be used to probe classical
as well as quantum turbulence, which helps making comparisons.

Panel A in Fig.~\ref{f:1} illustrates three different flow arrangements which
have enabled spectral measurements of velocity and vortex line density.
The configuration on the left is inspired by the historical experiment of Tabeling and
collaborators \cite{Tabeling} in which helium was driven by two counter-rotating propellers
attached to motors operating at cryogenic temperature (or at room temperature in more
recent experiments \cite{Schmoranzer-2009,SHREK}).
The configuration in the middle is the TOUPIE wind-tunnel \cite{Salort-EPL-2012}
which upgrades a smaller wind-tunnel \cite{Roche-2007}.
A 1-m-high column of liquid $^4$He pressurises hydrostatically the bottom
wind-tunnel section to allow operation in He-I up to velocities of
4 m/s %($\sqrt 2gh\simeq 4.5$ m/s)
without the occurrence of cavitation (prevented by the high thermal conductivity of
superfluid helium below the superfluid transition).
Without a dedicated pressurisation system,
the bubbles which would form in He~I
above $T_{\lambda}$ would prevent the comparison
of turbulence above and below the superfluid transition in the same apparatus.
The configuration at the right illustrates the TSF circulator, which
consists of
a pressurised helium loop cooled by a heat exchanger \cite{Salort-2010}.
All these flows are driven by the centrifugal force generated by propellers:
such forcing does not rely on viscous nor thermal effects and is therefore well
fitted to liquid helium irrespectively of its superfluid density fraction.

\begin{figure*}%%%%%%%%%%%%%%%  FRIG 2
\begin{tabular}{|c|c|c|}
  \hline
 A & B & C\\

\includegraphics[height=.29\textwidth]{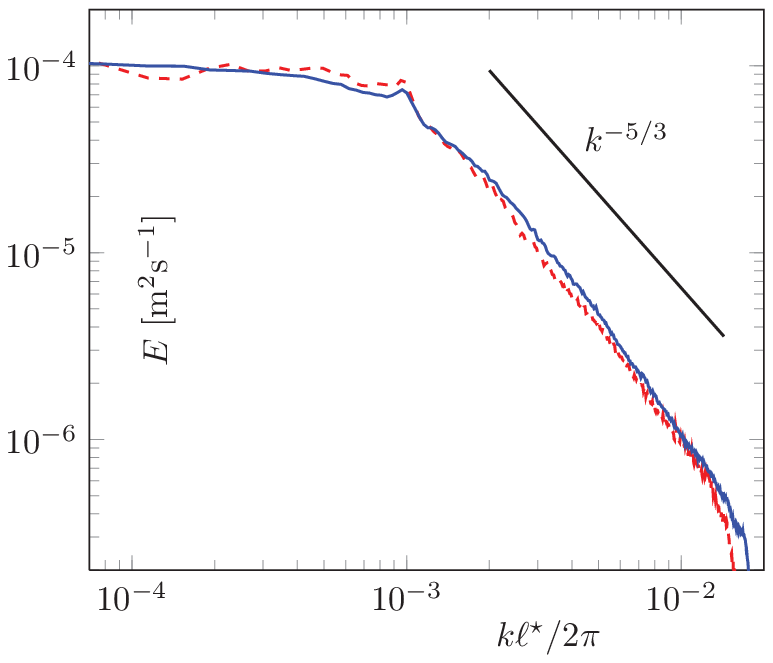}&
 \includegraphics[height=.3 \textwidth]{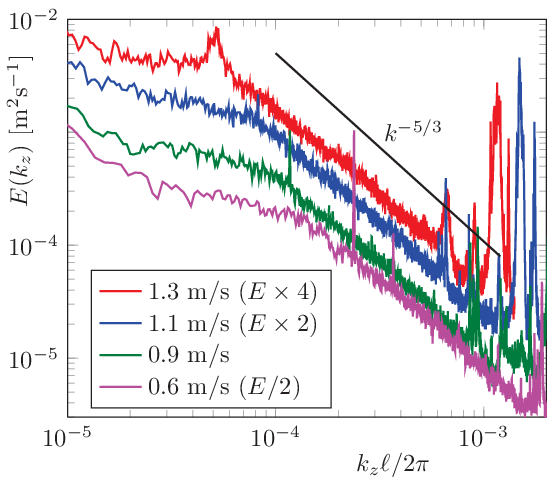}&
 \includegraphics[height=.24\textwidth]{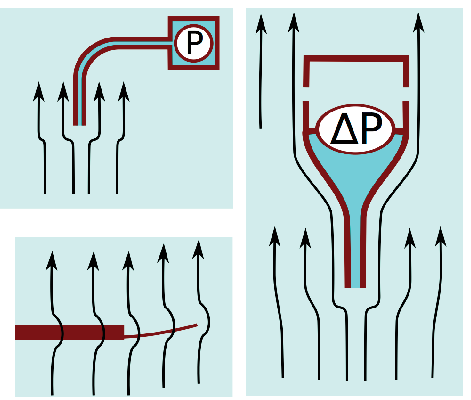}
  \\
 \hline
\end{tabular}
 \caption{\label{f:2}Color online.  Measured spectra of superfluid
turbulence in $^4$He. All cut-off at high $k$ or $f$ are caused
by the resolution of the probes.
Panel A: Energy spectrum measured in the TOUPIE wind-tunnel below the
superfluid transition (solid blue line, 1.56 K$< T_\lambda$)
and above $T_\lambda$ (dashed red line) \cite{Salort-EPL-2012}.
Above $T_\lambda$, the parameter $\ell^{\star}$ on the x-axis is chosen
such that the small peak separating the inertial scales plateau and the cascade
 matches the one estimated below $T_\lambda$
at $k_z \ell/2\pi=10^{-3}$
(this peak is associated with eddy shedding from the  cylinder
used upstream to generate turbulence).
Panel B: Energy spectra for different mean flow velocities for
$T = 1.55~{\rm K}$ in a smaller He-II wind tunnel.
An arbitrary vertical offset had been introduced
for clarity (see legend).
Panel C:
Local velocity probes used in superfluid $^4$He for measurements
of fluctuations
at sub-millimetre scales.
Stagnation pressure velocity probes without static pressure reference (top-left)
(\cite{Tabeling,Salort-2010} ), stagnation pressure velocity probe with a static pressure reference (right) (\cite{Salort-2010,Salort-EPL-2012}),
and cantilever-based velocity probe (bottom left) \cite{Salort-RSI-2012}.
%Vortex line density spectra measured near 1.6 K in the same He-II wind-tunnel
%as Panel B using miniature second-sound tweezers \cite{Roche-2007}.
%Some sharp noise peaks at high frequency have been
%removed from the datasets at the lowest velocities for clarity.
}
\end{figure*}

Experience shows that probing cryogenic flows is often more challenging than
producing the flow themselves, partly because dedicated probes often have to be
designed and manufactured for each experiment. This is all the more true if
good space and time resolutions are required to resolved the fluctuating
scales of superfluid turbulence.

Below $T_\lambda$, the most commonly-used local velocity probe is based
on the principle of the ``Pitot'' or ``Prandtl" tube
(sometimes called ``total head pressure'' tube), which is illustrated
in the top-left and right
sketches of Fig.~\ref{f:2}C. One end of a tube is inserted parallel
to the mean flow, while the other end is blocked by a pressure gauge.
The stagnation point which forms at the open end of the tube is associated with an
overpressure probed by the gauge. This stagnation-point overpressure $P$ is related
to the incoming flow velocity $V$ using Bernoulli relation $P \simeq \rho V^2 /2$.
In the arrangement depicted in the right sketch of Fig.~\ref{f:2}C,
the use of a differential pressure probe allows to remove the ``static''
pressure variation of the flow
associated with turbulent pressure fluctuations and acoustical background noise.
The operation of such stagnation-pressure probes below the critical temperature
and their limitations are discussed in Ref.~\cite{Salort-2010}.
In summary, the fluctuations $\delta P$ of $P$ are proportional to the
fluctuations $\delta V$ of $V$ up to the second order term in
$( \delta V / V ) ^2$ and the mean flow direction
has to be well defined (excessive angles of attack lead to measurement bias).
Pitot tubes achieving nearly 0.5-mm spatial resolution, and others with DC-4 kHz
bandwidth have been operated successfully.
At such scales and in the turbulent flows of interest, helium's two components
are expected to be locked together -as discussed later- and described by a single continuous fluid of total
density $\rho$. Therefore stagnation pressure probes determine
the common velocity of both fluid components.

The first experimental turbulent energy spectra below $T_{\lambda}$
were reported in 1998 \cite{Tabeling} using the set-up illustrated in
Fig.~\ref{f:1}A-left. Energy spectra at  2.08 K and 1.4 K were
found very similar to the spectrum measured in He~I
above the superfluid transition, at 2.3 K.
In the range of frequencies corresponding to the length scale of the
forcing scale and the smallest resolved length scale,
the measured spectrum
was compatible with KO--41.
The next published confirmation of  Kolmogorov's law came in
2010 \cite{Salort-2010}
from two independent wind-tunnels of the types depicted in the centre
and at the right-side of Fig.~\ref{f:1}A.
Measurements obtained with the first type of wind-tunnel are reproduced
in Fig.~\ref{f:2}B, which shows energy spectra
 at 1.6 K for various mean velocities of the flow.
We note that four decades separate the integral scale of the flow
($D \simeq$ 10 mm) and the intervortex scale $\ell \simeq 1 \rm \mu m$,
to be compared with the 1 mm effective resolution of the anemometer.
Measurements obtained with the second type of wind-tunnel explored grid
turbulence.
Although the signal-to-noise ratio was not as good, the choice of a
well-defined  flow
allowed to estimate independently both prefactors of Eq.~\eqref{K41}:
the Kolmogorov constant $C\Sb{K41}$ and the dissipation
rate $\varepsilon$.
Within accuracy (30\% for $C\Sb{K41}$), it was found that
both prefactors were similar above the superfluid
transition and below it in He-II at $T=2.0~\rm K$.
The energy spectrum shown
in Fig.~\ref{f:2}A has been recently obtained in the TOUPIE experiment
both above and below $T_\lambda$ in the far wake of a disc.
To normalise the x-axis of this plot, the mean intervortex distance $\ell$
in He-II was estimated
using the   relation $2{\ell} / {D}=$Re$_\kappa ^{-3/4}$~\cite{Salort-2011}
where ${\rm Re}_\kappa=D V/ \kappa$ is a Reynolds number defined using the root
mean square velocity from the anemometer, and the prefactor $2$ was
fitted to experimental and numerical data in the range
 $T\simeq 1.4-1.6\,K$.

The high signal-to-noise ratio of this dataset
allowed to check the validity of the $-4/5$
Karman-Howarth law \cite{Salort-EPL-2012}.
This law, sometimes described as the only exact relation known
in turbulence,
confirms that energy cascades from large to small scales without
dissipation within
the inertial range where the KO--41 scaling is observed.

Finally, it should be noticed that intermittency of velocity fluctuations
were partially explored in two experimental
studies \cite{Tabeling,SalortIntermi:ETC13_2011},
but no specific signature of superfluid turbulence was reported.
Both studies only explored the high and low temperature regimes
($\rho \sb s /\rho \sb n \simeq 0.29$ at $T=2.08\,K$ and
$\rho \sb s /\rho >0.85$ at $T \leq 1.56\,K$).

In all published energy spectra, the limited resolution of
the anemometer is responsible for the cut-off at high frequency/small scale.
Thus, the observed spectra reveal only ``half'' of the picture,
namely the integral scales and the upper half of the inertial scales.
To circumvent this limitation in resolution, a first approach is to
scale up the experiment (at given Reynolds number $Re_\kappa$)
so that all characteristic
flow scales are magnified and better resolved with existing probes.
This approach has been undertaken with the construction of a 78-cm diameter
He-II Von Karman flow in Grenoble \cite{SHREK} that is one order of
magnitude larger than the 1998's reference cell. Another approach is
to scale down the probes. For practical reasons it
is difficult to miniaturise much further stagnation pressure probes
without a significant decrease of their sensibility or time response.
New types of anemometers need to be invented.
One possibility arises from the recent development of fully
micro-machined anemometers based on the deflection of a silicon
cantilever (see the bottom left sketch of Fig.~\ref{f:2}C).
Preliminary spectral measurements with a resolution of $100 \rm \mu m$
have been recently reported in a He-II test facility \cite{Salort-RSI-2012}.

%%%%%%%%%%%%%%%%%%%%%%%%%%%%%%%%%%%%%%%%%%%%%%%%%%%%%%%%%%%%%%%%%

\section{\label{s:theory}   Equations of motion: three levels of description.}

In the absence of superfluid vortices, Landau's two-fluid
equations\cite{Donnelly} for the superfluid and normal fluid velocities
$\B u\sb s$ and $\B u\sb n$ account for all phenomena observed in He-II
at low velocities, for example
second sound and thermal counterflow. In the incompressible limit
(${\B \nabla} \cdot {\B u\sb s}=0$, ${\B \nabla}\cdot {\B u\sb n}=0$)
Landau's equations are:
\begin{subequations} \label{Landau} %%
\begin{eqnarray}\label{Landau1}
\Big[ (\partial \,\B u\sb s/\partial t)+ (\B u\sb s\cdot \B
\nabla)\B u\sb s \Big] &=&
-  \B \nabla p_s/ \rho_s,\\
%\end{equation}\begin{equation}
\label{Landau2}
\Big[ (\partial \,\B u\sb n/\partial t)+ (\B u\sb n\cdot \B
\nabla)\B u\sb n  \Big]&=& - \B \nabla p_n /\rho_n +  \nu_n \nabla^2 {\B u\sb n}
 \,,
\end{eqnarray}
\end{subequations}

\noindent
where $\nu_n=\mu/\rho_n$ is the kinematic
viscosity, and the efficient pressures $p_s$ and $p_n$ are defined by
${\B \nabla} p_s=(\rho_s/\rho){\B \nabla} p-\rho_s S {\B \nabla}T$ and
${\B \nabla} p_n=(\rho_n/\rho) {\B \nabla} p + \rho_s S {\B \nabla}T$.
On physical ground, Laudau argued that the superfluid is
irrotational.

The main difficulty in developing a theory of superfluid turbulence is the
lack of an established set of equations of motion for He-II in the presence of
superfluid vortices. We have only models at different
levels of description.

\subsection{\label{4A} First level}
At the first, most microscopic level of description, we must account for
phenomena at the length scale of the
superfluid vortex core, $R \approx \xi$.
Monte Carlo models of the vortex core
\cite{Reatto}, although
realistic, are not suitable for the study of the dynamics and turbulent motion.
A practical model of a pure superfluid is the
Gross-Pitaevskii Equation (GPE) for a weakly-interacting Bose gas~\cite{Annett}:
\begin{equation} \label{GPE}
i \hbar \frac{\partial \Psi}{\partial t}=
-\frac{\hbar^2}{2 M} \nabla^2 \Psi + V_0 \vert \Psi \vert^2 \Psi -E_0 \Psi\,,
\end{equation}
\noindent
where $\Psi(\B r,t)$ is the complex condensate's wave function,
$V_0$ the strength of the interaction between bosons,  $E_0$ the
chemical potential and $M$ the boson mass. The condensate's
density $\tilde{\rho_s}$ and velocity $\tilde{\B v\sb s} $
are related to $\Psi=\vert \Psi \vert \exp(i \Theta)$ via the Madelung
transformation
$
\tilde{\rho_s}=M \vert \Psi \vert^2\,,
\quad \tilde{\B u\sb s}= \hbar \B \nabla \Theta/ M$,
which confirms Landau's intuition that the superfluid is
irrotational.
It can be shown that, at length scales $R \gg \xi=\hbar/\sqrt{2 M E_0}$,
the GPE reduces
to the classical continuity equation and the (compressible) Euler equation.
It must be stressed that,
although the GPE accounts for quantum vortices,
finite vortex core size (of the order of $\xi$), vortex nucleation,
vortex reconnections, sound emission by accelerating vortices and
Kelvin waves, it is only a qualitative model of the superfluid component.
He-II is a liquid, not a weakly-interacting gas, and the
condensate is only a fraction of the superfluid density $\rho_s$.
No adjustment of $V_0$ and $E_0$ can fit both the sound speed and the
vortex core radius, and the dispersion relation of the uniform solution
of Eq.~\eqref{GPE} lacks
the roton's minimum which is characteristic of He-II \cite{Donnelly}.
%\cite{Pomeau:1993p3780}.
 This is why, strictly
speaking, we cannot identify $\tilde{ \rho_s}$ with $\rho_s$ and
$\tilde{ \B u\sb s}$ with $\B u\sb s$. Nevertheless, when solved
numerically, the GPE is a useful model of superfluid turbulence at
low $T$ where the normal fluid fraction vanishes,
and yields results which can be compared to experiments, as
we shall see.

\subsection{\label{4B} Second level}
Far away from the vortex core at length scales larger
than $\xi$ and in the zero Mach number limit, the GPE describes
incompressible Euler dynamics. This is the level of description of a second
practical model, the Vortex Filament Model (VFM) of Schwarz\cite{Schwarz1988}.
At
this level (length scales $R \gg \xi$)
we ignore the nature of the vortex core but distinguish individual
vortex lines, which we describe as oriented space curves $\B s(\xi,t)$ of
infinitesimal thickness and circulation $\kappa$, parametrised
by arc length $\zeta$.  Their time evolution is determined by Schwarz's
equation
\begin{subequations}\label{VFM}
\begin{equation}\label{VFMa}
\frac{d{\bf s}}{dt}=\B u \sb {si}+ \B w\,, \quad \B u \sb {si}(\B s)=
\frac{\kappa}{4 \pi} \oint_{\cal L} \frac{({\bf s_1}-{\bf s})
\times d{\bf s_1}}{|\B s_1-\B s|^3},
\end{equation}
where the self-induced velocity $\B u \sb {si}$ is given by the
Biot-Savart law \cite{Saffman},
and the line integral extends over the vortex configuration. At nonzero
temperatures the
term $\B w$ accounts for the friction between the vortex lines
and the normal fluid\cite{BDV1982}:
\begin{equation}\label{VFMcoupling}
\B w=\alpha {\bf s}' \times \B u\sb{ns}
- \alpha' {\bf s}' \times [\B s' \times \B u\sb{ns}]\,,
\quad  \B u \sb {ns}= {\B u \sb n}-{\B u \sb {si}}\,,
\end{equation}
\end{subequations}
where
$\B s' = d \B s / d\zeta$ is the unit tangent at $\B s$,
and $\alpha$, $\alpha'$ are known \cite{DB}
temperature-dependent friction coefficients. In the very low
temperature limit ($T \to 0$),
$\alpha$ and $\alpha'$ become negligible\cite{DB},
and we recover the classical
result that each point of the vortex
line is swept by the velocity field produced by the entire vortex
configuration.

In numerical simulations based on the VFM, vortex lines are discretized
in a Lagrangian fashion, Biot-Savart integrals are
desingularised using the vortex core radius $\xi$,
and reconnections are additional artificial
\emph{ad-hoc} procedures that change the way pairs of discretization
points are connected. Reconnection criteria
are described and discussed in Ref.~\cite{Baggaley-tree,Baggaley-recon};
Ref.~\cite{Zuccher} compares GPE and VFM reconnections with each other
and with experiments. Simulations at large values of vortex line density
are performed
using a tree algorithm\cite{Baggaley-tree} which speeds up the
evaluations of Biot-Savart integrals from $N^2$ to $N\log{N}$ where $N$ is
the number of discretization points.
The major drawback of the VFM is that the normal fluid $\B u \sb n$ is
imposed (either laminar or turbulent), therefore the back-reaction of the
vortex lines on $\B u \sb n$ is not taken into account.
The reason is the computational difficulty: a self-consistent
simulation would require
the simultaneous integration in time of Eq. ~\eqref{VFM} for
the superfluid, and of a Navier-Stokes equation for the normal
fluid (implemented with suitable friction forcing at vortex lines
singularities).
Such self-consistent simulations were carried out only for a single
vortex ring \cite{Kivotides-ring} and for the initial growth of a
vortex cloud \cite{Kivotides-2011}.
This limitation is
likely to be particularly important at low  and intermediate
temperatures (at high temperatures
the normal fluid contains most of the kinetic energy, so it is less likely
to be affected by the vortices).

\subsection{\label{4C} Third level}
At a third level of description we do not distinguish individual vortex
lines any longer, but rather consider fluid parcels which contain
a continuum of vortices. At these length scales $R \gg \ell$
we seek to generalise
Landau's equations \eqref{Landau} to the presence of vortices.
In laminar flows the vortex lines (although curved) remain locally
parallel to each other, so it is possible to define the components of
a macroscopic vorticity
field $\bom_s$ by taking a small volume larger than
$\ell$ and considering the superfluid circulation in the
planes parallel to the Cartesian directions (alternatively,
the sum of the oriented vortex lengths in each Cartesian direction).
We obtain the so-called Hall-Vinen (or HVBK) ``coarse-grained" equations
\cite{Hall-Vinen,Hills-Roberts}:
\begin{subequations} \label{NSE} %%
\begin{eqnarray}\label{NSEa}
&&\hskip - 0.7cm  \Big[\frac{\partial \,\B u\sb s}{\partial t}+ (\B u\sb s\cdot \B
\nabla)\B u\sb s \Big] =
- \frac{1}{\rho_s}{\B \nabla p_s}-\B f \sb {ns},
%\end{equation}\begin{equation}
\\ \label{NSEb}
&&\hskip - 0.7cm \Big[\frac{\partial \,\B u\sb n}{\partial t}+ (\B u\sb n\cdot \B
\nabla)\B u\sb n  \Big]=
-\frac{1}{\rho_n} {\B \nabla p_n}
+  \nu_n \nabla^2 {\B u\sb n}+\frac{\rho_s}{\rho} \B f \sb {ns},
%\end{equation}\begin{equation}
\\ \label{NSEc}
&&\hskip - 0.7cm \B f \sb {ns}=\alpha {\hat \bom_s} \times (\bom_s \times  \B u \sb {ns})
+\alpha' {\hat \bom_s} \times \B u \sb {ns},
\end{eqnarray}
\end{subequations}

\noindent
where $\bom_s=\nabla \times \B u\sb s$,
${\hat \bom}_s=\bom_s/\vert \bom_s \vert$ and $\B f \sb {ns}$ is the
mutual friction force. These equations have been used with success to
predict the Taylor-Couette flow of He-II, its stability
\cite{Barenghi-Couette} and the
weakly nonlinear regime \cite{Henderson}. In these flows, the vortex
lines are fully polarised and aligned in the same direction, and
their density and orientation may change locally and
vary as a function of position (on length scales larger than $\ell$).

The difficulty with applying the HVBK equations to turbulence
is that in
turbulent flows the vortex lines tend to be
randomly oriented with respect to each other, so
the components of $\B s'$ cancel out to zero (partially or totally),
resulting in local vortex length (hence energy dissipation) without any
effective superfluid vorticity. In this case, the HVBK equations
may become a poor approximation
and underestimate the mutual friction coupling. Nevertheless, they are
a useful model of large scale superfluid motion with
characteristic scale $R\gg \ell$, particularly because (unlike the VFM)
they are dynamically self-consistent (normal fluid
and superfluid affect each other).
We must also keep in mind that Eq.~\eqref{NSE} do not
have physical meaning at length scales smaller than $\ell$.
In the next sections we shall describe
numerical simulations of equations~\eqref{NSE} as well as shell models
and theoretical models based on these equations. In some models
the mutual friction force is simplified to
$\B f \sb {ns}=-\alpha \kappa {\cal L} \B u \sb {ns}$ where
${\cal L}=1/\ell^2$.

Numerical simulations in the framework of all three approaches~\eqref{GPE},
\eqref{VFM} and \eqref{NSE} are shown in Figs.~\ref{f:3}.
They clearly show KO--41 scaling,
in agreement with the experimental results
shown in Figs.~\ref{f:2}.
Details of this simulations will be described below.
%%%%%%%%%%%%%%%%%%%%%%%%%%%%%%%%%%%%%%%%%%%%%%%%%%%%%%%%%%%%%%%%%

\section{\label{s:num}   Numerical experiments.}

Since the pioneering work of Schwarz \cite{Schwarz1988}, numerical
experiments have played an important role,
allowing the exploration of the consequences of limited
sets of physical assumptions in a controlled way, and providing
some flow visualization.
%\red{For years the study of helium hydrodynamics has suffered from the lack of direct techniques of visualisation (compared to what is available at room temperature in ordinary fluids);} fortunately new visualisation tools have recently become available \cite{Nugzar2013,VanSciver2009,Fonda,Guo,LaMantia:JFM2013}.

 \subsection{\label{5A} The GPE}
Numerical simulations of the GPE in a three-dimensional
periodic box have been performed
for decaying turbulence \cite{Nore} following an imposed arbitrary
initial condition, and for forced turbulence \cite{Kobayashi,Sasa-2011}.
Besides
vortex lines, the GPE describes compressible motions and sound
propagation; therefore, in order to analyse turbulent vortex lines,
it is necessary to extract from the total energy of the system
(which is conserved during the evolution) its incompressible kinetic energy
part whose spectrum is relevant to our discussion.
To reach a steady state, large-scale external forcing and small-scale
damping was added to the GPE~\cite{Sasa-2011}. The resulting turbulent
energy spectrum,  shown in Fig.~\ref{f:3}C, agrees with the
KO--41 scaling (shown by cyan dot-dashed line),
and demonstrates bottleneck energy accumulation near the intervortex
scale at zero temperature predicted  earlier in~\cite{LNR-2}
and discussed  in Sect.~8A.
The KO--41 scaling observed in GPE simulations was found to be
consistent with the VFM at zero temperature~\cite{Araki,Baggaley-structures}
and has also been observed when modelling a trapped atomic Bose--Einstein
condensate \cite{Tsubota-PLTP}.

The GPE can be extended to take into account finite temperature effects.
Different models have been proposed
\cite{Proukakis:2008,Brachet:CRPhys2012,Allen2013,Krstulovic:PRB2011}.

%%%%%%%%%%%%%%%%%%%%%%%%%%%%%%%%%%%%%%%%%%%%%%%%%%%%%%%%%%%%%%%%%

 \subsection{\label{5B} The VFM}

Most VFM calculations have been performed in a cubic box of size $D$
with periodic boundary conditions\footnote{De Waele and collaborators\cite{Aarts} used solid boundary
conditions \cite{Aarts} and investigated flat and a
parabolic normal fluid profiles, an issue which is still open.}.
In all relevant experiments we expect that the
normal fluid is turbulent because its Reynolds number $Re=D V_n/\nu_n$
is large (where $V_n$ the root mean square
normal fluid velocity).
Recent studies thus assumed the form
\cite{Sherwin-2012,Kivotides-2002,Laurie-2012}\\
% \begin{equation}
$$
 {\bf u}_n({\bf s},t)=\sum_{m=1}^{M}({\bf A}_m \times {\bf k}_m \cos{\phi_m}
+{\bf B}_m \times {\bf k}_m \sin{\phi_m}),
%\label{eq:KS}\end{equation}
$$
where $\phi_m={\bf k}_m \cdot {\bf s} + f_m t$, ${\bf k}_m$ and
$f_m=\sqrt{k^3_m E(k_m)}$ are  wave vectors and angular frequencies.
The random parameters ${\bf A}_m$,  ${\bf B}_m$  and ${\bf k}_m$ are chosen
so that the normal fluid's energy spectrum obeys  KO--41 scaling
$E(k_m)\propto k_m^{-5/3}$ in the inertial range $k_1<k< k_M$.
This synthetic turbulent flow~\cite{Osborne-2006} is solenoidal, time-dependent,
and compares well with Lagrangian statistics obtained in experiments and
direct numerical simulations of the Navier-Stokes equation.
Other VFM models included normal-fluid turbulence
generated by the Navier--Stokes equation \cite{Morris}
and a vortex-tube model
\cite{Kivotides-2006}, but,
due to limited computational resources,
only a snapshot of the normal fluid, frozen in time, was used to drive
the superfluid vortices.
 % Typical numerical experiments with the VFM start from a small number of vortex loops
% as initial condition;

In all numerical experiments, after a transient from some initial condition,
a statistical steady state of
superfluid turbulence is achieved, in the form of a vortex tangle
(see Fig.~\ref{f:1}-B) in which the vortex line density
$\cal{L}$$(t)$
fluctuates about an average value independent of the initial condition.

Recent analytical~\cite{LNR-1} and numerical
studies\cite{Sherwin-2012,Laurie-2012} of the geometry
of the vortex tangle
reveal that the vortices are not randomly distributed,
but there is a tendency
to locally form bundles of co-rotating vortices, which keep forming,
vanish and reform somewhere else. These bundles
associate with the Kolmogorov spectrum: if turbulence is driven
by a uniform normal fluid (as in the original work of Schwarz\cite{Schwarz1988}
recently verified in Ref.~\cite{Adachi}), there are
nor Kolmogorov scaling nor bundles.
Laurie et al.\cite{Laurie-2012} decomposed the vortex tangle
in a polarised part (of density $L_{\parallel}$) and a random part
(of density $L_{\times}$), as argued by Roche \& Barenghi \cite{Roche-2008},
and discovered that $L_{\parallel}$ is responsible for the $k^{-5/3}$ scaling
of the energy spectrum, and $L_{\times}$ for the $f^{-5/3}$ scaling of the
vortex line density fluctuations, as suggested in Ref.\cite{Roche-2007}.

\subsection{\label{5C} The HVBK equations}
From a computational viewpoint, the HVBK equations
are similar to
the Navier-Stokes equation \eqref{NavierStokes}.
Not surprisingly, standard methods of classical turbulence have
been adapted to the HVBK equations, e.g. Large Eddy
Simulations \cite{Merahi:2006},  Direct Numerical
Simulations \cite{Roche2fluidCascade:EPL2009,Salort-2011}
and
Eddy Damped QuasiNormal Markovian simulations \cite{Tchoufag:POF2010}.

%\magenta{\textbullet~}i
The HVBK equations are ideal to study the coupled dynamics of
superfluid and normal fluid in the limit of
intense turbulence at finite temperature.
Indeed, by ignoring the details of individual vortices and their fast dynamics,
HVBK simulations do not suffer as much as VFM and GPE simulations
from the wide separation of space and time scales which characterize
superfluid turbulence. Moreover, well optimized numerical solvers have
been developed for Navier-Stokes turbulence and they can be easily
adapted to the HVBK model.
Thus, simulations over a wide temperature range
($1.44<T<2.157~{\rm K}$
corresponding to $0.1\le \rho_n / \rho_s \le 10$) evidence a strong
locking of superfluid and normal fluid ($\B u\sb s \approx \B u\sb n$)
at large scales, over
one decade of inertial range (\cite{Roche2fluidCascade:EPL2009}).
In particular, it was found that even if one single single fluid
is forced at large scale (the dominant one), both fluids still get locked
very efficiently.

Fig.~\ref{f:3}A illustrates velocity spectra generated by direct
numerical simulation of the HVBK equations, while
the red and blue solid lines of Fig.~\ref{f:3}B
show spectra obtained using a shell model  of the same equations
(see paragraph at the end of the section). In both case, a clear
$k^{-5/3}$ spectrum is found for both fluid components, at  all
temperature and large scale.

Information about the quantization of vortex lines is lost in the coarse
graining procedure which leads to Eqs.~\eqref{NSE}.
As discussed in
Sect.~8B, a quantum constrain can be re-introduced in this
model by truncating  superfluid phase space for
$| \B k | \leq \ell ^{-1}$,
causing an upward trend of the low temperature
velocity spectrum of Fig.~\ref{f:3}A which can be
interpreted as partial thermalization
of superfluid excitations.
This procedure also leads to the
prediction ${\cal L} D^2=4 \mbox{Re}^{3/2}$~\cite{Salort-2011}
which is consistent with experiments and  allows to identify
the spectrum of $\cal L$$(r) / \kappa$ with
the spectrum of the scalar field $| \omega_s (r) |$.
It is found that this
spectrum is temperature--dependent in the inertial range
with a flat part at high temperature (reminiscent of the corresponding
spectrum of the magnitude of the vorticity in classical turbulence)
%\cite{Ishihara2003,Brachet:1991})
which contrasts the $k^{-5/3}$ decreases at low temperature
(consistent with experiments \cite{Roche-2007}).

%\magenta{\textbullet~
Essential simplification of the HVBK Eqs.~\eqref{NSE} can be achieved
with the shell-model
approximation\cite{Wacks-2011,Boue:PRB2012,Boue:PRL2013}.
The complex shell velocities $u_m \sp{s} (k_m)$ and
$u_m \sp{n} (k_m)$ mimic the statistical behaviour of the
Fourier components of the turbulent superfluid
and normal fluid velocities at wavenumber $k$.
The resulting ordinary differential Eqs. for $u_m\sp{n,s}$
capture important aspects of the HVBK
Eqs.~\eqref{NSE}. Because of the geometrical spacing of the shells
($k_m = 2^m k_0 $), this approach allows more decades
of $k$-space than Eqns.~\eqref{NSE}
(see Fig.~\ref{fig:interim} with eight decades
in $k$-space~\cite{Boue:PRL2013}).
This extended inertial range allows detailed comparison of
intermittency effects
in superfluid turbulence and  classical turbulence
(see Sec.~6C).

%%%%%%%%%%%%%%%%%%%%%%%%%%%%%%%%%%%%%%%%%%%%%%%%%%%%%%%%%%%%%%%%%

\section{\label{ss:HD}
  Models: the hydrodynamic range
}
In this section we discuss theoretical models of
large scale (eddy dominated) motions at wavenumbers $k \ll k_{\ell}$
which are important at all temperatures from 0 to $T_\lambda$. These motions
can be tackled using the hydrodynamic HVBK Eqs.~\eqref{NSE}, thus
generalising what we know about classical turbulence. We shall start
by considering the simpler case of $^3$He (Sect.~\ref{6A}),
in which there is only one turbulent fluid, then move to more difficult
case of $^4$He (Sect.~\ref{6B}) in which there are two coupled turbulent fluids,
and finally discuss intermittency (Sect.~\ref{6C}).

%Secondly (Sect.~6D), we
%consider small scale motions with $k \gg \ell$, dominated by Kelvin waves,
%which are particularly important at very low temperatures below $1\,$K.
%The difficult \blue{cross-over} region $k \approx \ell$, where both types of
%motion are important, will be tackled in Sect.~7A. Sect.~7B describes an
%alternative model which has been proposed\cite{Salort-2011},
%in which the discrete nature of superfluid vorticity
%affects the spectrum even at $k<k_{\ell}$}.

 \subsection{\label{6A}
  One turbulent fluid: $^3$He}
The viscosity of $^3$He is so large that, in all $^3$He turbulence
experiments, we expect the normal fluid to be at rest ($\B u\sb n=0$)
or in solid body rotation in rotating cryostat (in which case our argument requires
a slight modification). Liquid $^3$He thus provides us with
a simpler turbulence problem (superfluid turbulence in the presence
of linear friction against a stationary normal fluid) than $^4$He
(superfluid turbulence in the presence of normal fluid turbulence).
At scales $R \gg \ell$, we expect
Eq.~\eqref{NSEa} to be valid, provided we add a suitable model for the friction.
Following Ref.~\cite{LNV}, we approximate $\B f \sb {ns}$ as\\
 \begin{equation}\label{NSEf}
 \B f \sb {ns} = -\alpha \kappa {\cal L} {\B u\sb {ns}}=-\Gamma {\B u\sb {ns}}\,, \
 \langle \vert \bom_s \vert^2 \rangle
\approx 2\int_{k_0}^{1/\ell} k^2 E(k) d k\,,
 \end{equation}%%
with ${\B u \sb n}={\bf 0}$.
Here  $\Gamma=\alpha \kappa \o\Sb T$,   $\o\Sb T \equiv \sqrt{ \langle \vert \bom_s \vert^2 \rangle}$
  is   the characteristic ``turbulent"
superfluid vorticity, estimated via the spectrum $E(k)$.
$k_0$ is the energy pumping scale.
Using the differential approximation~\eqref{Leith-n} for the energy
spectrum, the continuity Eq.~\eqref{EnB} in the stationary case becomes
\begin{equation}\label{balA}
\frac{1}{8}\frac{d}{d k}\Big [ \sqrt{k^{11}E\sb s(k)}\frac{d}{dk}\frac {E\sb s(k)}{k^2}\Big ] + \Gamma E\sb s (k)=0.
\end{equation}

Analytical solutions of Eq.~\eqref{balA}
are in good agreement \cite{LNV,He3spectra}
with the results of numerical simulation of
the shell model to the hydrodynamic Eq.~\eqref{NSEa}, providing us with
quasi-qualitative description of turbulent
energy spectra in $^3$He over a wide region of temperatures.

\subsection{\label{6B} Two coupled turbulent fluids: $^4$He}

%\green{\Fbox{In the section below I have replaced $\nu$ with $\nu_n$, because in the
%absence of mutual friction coupling, Eq.14b must become the Navier Stokes equation
%and $\nu_n \to \nu$
%}}
In $^4$He, we have to account for motion of the normal component,
which has very low viscosity
and is turbulent in the relevant experiments.
Eqs.~\eqref{NSEa}, \eqref{NSEb} and \eqref{NSEf}
(now with $\B u\sb n \neq 0$)
result in a
system of energy balance equations for superfluid and normal fluid
energy spectra $E\sb s(k)$ and $E\sb n(k)$ that generalises
Eq.~\eqref{balA}~\cite{LNS}:
\begin{subequations}\label{2-bal}\begin{eqnarray}\label{2-balA}
 && \hskip -0.15cm \frac{d \varepsilon\sb s (k)}{d k}+ \Gamma \big[E\sb s (k)-E\sb  {ns} (k)]=0\,, \\ \label {2-balB}
 && \hskip -0.5cm \frac{d \varepsilon\sb n (k)}{d k}+ \frac{\rho\sb s}{\rho\sb n} \Gamma  \big[E\sb n (k)-E\sb  {ns} (k)]=
-2 \nu_n k^2 E\sb n (k)\ .~~~~~~~
\end{eqnarray}
Here superfluid and normal fluid energy fluxes $\varepsilon\sb s (k)$
and  $\varepsilon\sb n (k)$ can be expressed via
$E\sb s(k)$ and $E\sb n(k)$ by differential closure~\eqref{Leith-n}.
The cross-correlation function $E\sb  {ns} (k)$ is normalized
such that
$\int E\sb  {ns} (k) dk= \langle \B u\sb s \cdot \B u\sb n  \rangle $.
If, at given $k$, superfluid and normal fluid eddies
are fully correlated (locked),
then $E\sb  {ns} (k)=E\sb  {s} (k)=E\sb  {n} (k)$.
If they are statistically independent (unlocked),  then $E\sb  {ns} (k)=0$.
The  closure equation for $E\sb  {ns} (k)$  was proposed
in~Ref.~\cite{LNS}:
\begin{eqnarray} \label {2-balC}   E\sb{sn}(k) &=&
\frac { \rho \sb s    E\sb s(k)+
\rho \sb n  E\sb n(k) }{  \rho\,  [ 1+K(k)]}, \\
 K(k)&\equiv&\frac{\rho\sb n
[\nu_n  k^2 +\gamma \sb n(k)+\gamma \sb s(k)]}{ \rho\, \alpha\omega \sb T}\ .
\end{eqnarray}
\end{subequations}
Here  $  \gamma \sb n(k) \simeq  k\sqrt { k E\sb n (k)}$ and $  \gamma \sb s(k)\simeq   k\sqrt {k  E\sb s (k)} $ are characteristic interaction frequencies (or turnover frequencies) of eddies in the normal and superfluid components. They are related to the well known effective turbulent viscosity $\nu\Sb T$
by $\nu\Sb T k^2=  \gamma (k)$.

 For large mutual friction or/and small $k$, when $K(k)\ll 1$ and
can be neglected,  Eq.~\eqref{2-balC} has a physically motivated solution
$$
E_{sn}(k)=E_{s}(k)=E_{n}(k)$$
 corresponding to full locking
$\B u\sb n(\B r, t)=\B u\sb s(\B r, t)$.
In this case the sum of Eq.~\eqref{NSEa} (multiplied by $\rho\sb s$)
and Eq.~\eqref{NSEb} (multiplied by $\rho\sb n$) yields the Navier-Stokes
equation with effective viscosity $\widetilde \nu=\nu \rho\sb n/\rho$.
Thus, in this region of $k$, one expects classical behaviour of
hydrodynamic turbulence with KO--41 scaling~\eqref{K41}
(up to intermittency corrections discussed in Sec.~6C).

For small mutual friction or/end large $k$, when $K(k)\gg 1$,
Eqs.~\eqref{2-balC} gives~\cite{LNS}
$$
E\sb{sn}(k)\ll \sqrt {E\sb n (k) E\sb s(k)}\,,%%
$$
i.e. full decorrelation of superfluid and normal fluid velocities.
In this case normal component will have KO--41scaling~\eqref{K41},
$$E\sb n(k)\simeq \varepsilon\sb n ^{2/3}k^{-5/3}\,,$$
up to the Kolmogorov micro-scale $k_\eta$ that can be found from the
condition $\nu_n k_\eta^2\simeq \gamma\sb n (k_\eta)$, giving the
well known estimate
 $$
 k_\eta \simeq \varepsilon\sb n ^{1/4}/ \nu_n ^{3/4}\ .%%
 $$
Simultaneously, the superfluid spectrum also obeys the same KO--41
scaling, $E\sb s(k)\simeq \varepsilon\sb s ^{2/3}k^{-5/3}$.
Moreover, since at small $k$ the two fluid components are locked,
we expect that $\varepsilon\sb s\approx \varepsilon\sb n$.
Assuming that the $5/3$ scaling is valid up to wavenumber $k_\ell = 1/\ell$,
we estimate that
 $$
 k_\ell \simeq  \varepsilon_s ^{1/4}/ \kappa ^{3/4}\,, %%
 $$
which is similar to the estimate for $k_\eta$ with the replacement of
$\nu_n$ with $\kappa$. In He-II, the
numerical values of $\nu\sb n$ and $\kappa$ are similar,
%\red{($\nu\sb n/\kappa \approx 1$ to $0.2$ from $1.45~K$ to $2.15~K})
%($\nu  \sim \kappa \rho / (6 \rho\sb n)$),
thus we conclude
that the viscous cutoff $k_\eta$ for the normal component and the
quantum cutoff $k_\ell$ for the superfluid component are close to each
other.

At a given temperature, the decorrelation crossover wave vector $k_*$
between the two regimes described above can be found from the
condition $K(k_*)=1$ using the estimate
$$
\gamma\sb n(k_*)\simeq \gamma\sb s(k_*) \simeq
\varepsilon \sb s^{1/3}k_*^{2/3} \gtrsim \nu k^2_*%%
$$
with $\omega\Sb T\simeq \varepsilon \sb s^{1/3}k^{2/3}_\ell $. We obtain
$k_*\ell \simeq  (\rho\sb n/ \alpha\rho)^{3/2}$.
 The quantity $\alpha\rho/\rho\sb n$
varies between 1.2 and 0.5 \cite{DB} in the temperature range
$0.68 < (T/T_\lambda) < 0.99$ where the motion of the normal fluid
is important.
We conclude that  $k_*\simeq k_\ell$, which means that
normal fluid and superfluid eddies
are practically locked over the entire inertial interval.
Nevertheless,  dissipation due to mutual friction
cannot be completely ignored,   leading to intermittency enhancement
 described next.

%\blue{
%Section 7B presents an alternative approach \cite{Salort-2011} which doesn't not assume a $-5/3$ scaling
%down to scale $\ell$ and explores the possibility that the discrete nature of superfluid
%vorticity affects the spectrum even at wavenumbers $k < k_{\ell}$,
%over a range of scales where both fluids are unlocked.
%}

\subsection{\label{6C} Intermittency enhancement in $^4$He}

The first numerical study of intermittent exponents
\cite{SalortIntermi:ETC13_2011}
did not find any intermittent effect peculiar to superfluid
turbulence neither at low
temperature ($T\simeq 0.5 T_\lambda$, $\rho \sb s / \rho \sb n=40$)
nor at and high temperature
($T\simeq 0.99 T_\lambda$, $\rho \sb s / \rho \sb n=0.1$),
in agreement with experiments
performed at the same temperatures (see Sect.~\ref{s:exp}).

\begin{figure}
\centerline{\includegraphics[width=0.5\textwidth]{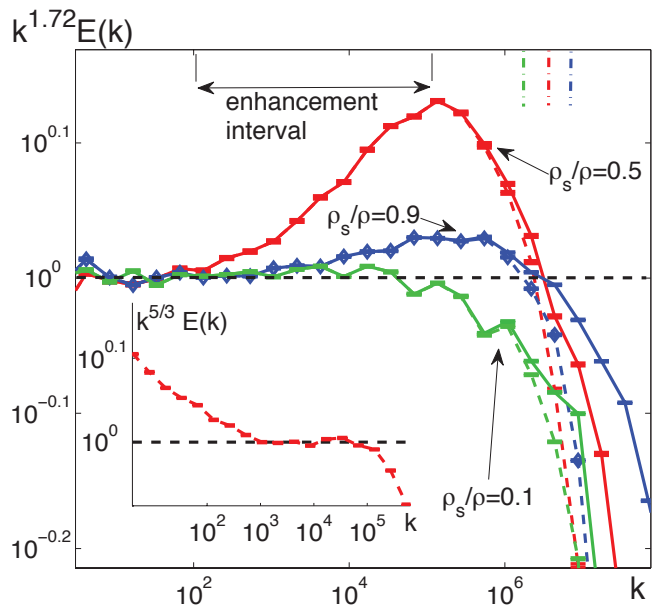}}
\caption{\label{fig:interim}
Superfluid (solid lines) and normal fluid (dash lines)
compensated energy spectra  $k^{1.72}E(k)$;
the compensation factor is the classical energy spectrum with
intermittency correction.
Inset: $k^{5/3}E(k)$  for $T=0.9\,T_\lambda$.
Shell model simulation of the HVBK model at $T/T_\lambda = 0.99$ K (green),
0.9 (red) and 0.9  (blue), corresponding to
$\rho\sb s/\rho=0.1\,, \ 0.5\,,$ and 0.9 respectively~\cite{Boue:PRL2013}.
The vertical dash lines indicate $k_\ell\equiv 1/\ell$.}
\end{figure}

Recent shell model simulations \cite{Boue:PRL2013} with eight
decades of $k$-space allowed detailed comparison of
classical and superfluid turbulent statistics in the intermediate
temperature range corresponding to $\rho \sb s \approx \rho \sb n$.
The results were the following.
For $T$ slightly below
$T_\lambda$, when $\rho \sb s /\rho \sb n\ll 1$,
the statistics of turbulent superfluid $^4$He  appeared similar
to that of classical fluids, because the superfluid component can be
neglected, see green lines in
Fig.~\ref{fig:interim} with $\rho\sb s/ \rho=0.1$.
The same result applies to   $T\ll T_\lambda$
($\rho\sb n \ll \rho \sb s$), as expected due to the inconsequential
role played by the normal component, see blue lines
with $\rho\sb s/ \rho=0.9$.
In agreement with the previous study
the intermittent scaling exponents appeared the same in classical and
low-temperature superfluid turbulence (indeed the nonlinear
structure of the equation for the superfluid
component is the same as of Euler equation, and dissipative
mechanisms are  irrelevant.)

A difference between classical and superfluid intermittent
behaviour in a wide (up to three decades) interval of scales
was found in the range $0.8 T_{\lambda} < T < 0.9 T_{\lambda}$
($\rho\sb s \approx \rho \sb n$), as shown by red lines in
Fig.~\ref{fig:interim} with $\rho\sb s/ \rho=0.5$.
%to diminishes from intermittency  value 1.72
%down to  $0.67$, close to the K041 value 5/3.  [NOT CLEAR]
The  exponents of higher order correlation functions
also deviate further from the KO--41 values. What is predicted is thus
an enhancement of intermittency
in superfluid turbulence compared to the classical turbulence.

\section{\label{VII}
  Models: the Kelvin wave range }

Now we come to the more complicated and more
intensively discussed aspect of the superfluid energy spectrum:
what happens for $k\ell\gg  1 $, where the
quantisation of vortex lines becomes important.
This range acquires great importance
at low temperatures, typically below 1 K in $^4$He, and is relevant
to turbulence decay experiments.
Here we shall describe only the basic ideas, avoiding the most
debated details.

For $k\ell\gg 1$ we neglect the interaction between separate
vortex lines (besides the small regions around vortex reconnection events,
which will be discussed later). Under this reasonable assumption,
at large $k$ superfluid turbulence can be considered as a system of
Kelvin waves (helix-like deformation of vortex lines) with different
wavevectors interacting with each other on the same vortex.
The prediction that this interaction results in turbulent energy transfer
toward large $k$~\cite{Svi95} was confirmed by numerical simulations
in which energy was pumped into Kelvin waves at
intervortex scales by vortex reconnections~\cite{Carlo2} or simply by
exciting the vortex lines~\cite{Tsubota3}.
The first analytical theory of Kelvin wave turbulence (propagating
along a straight vortex line and in the limit of small amplitude compared
to wavelength) was proposed by Kozik and Svistunov~\cite{KS}  (KS),
who showed that the leading interaction is a six-wave scattering process
(three incoming waves and three outgoing waves).
Under the additional assumption of locality of the interaction
(that only compatible wave-vectors contribute to most of the energy transfer)
KS found that
(using the same normalisation of other hydrodynamic spectra such as
Eqs.~\eqref{K41})
%and \eqref{He3A})
the energy spectrum of Kelvin waves is \\ \vskip 0cm
 $\displaystyle
	 E\Sp{KS}\Sb{KW}(k)\simeq \Lambda \,  \varepsilon^{1/5}\Sb{KW}\kappa^{7/5}\, \ell^{-8/5}k^{-7/5}, \ \qquad \mbox{(KS spectrum).} $ \\ \vskip 0cm
\noindent Here $\Lambda\equiv \ln (\ell/\xi)\simeq 12$ or $15$ in
typical $^4$He and $^3$He experiments, and $\varepsilon\Sb{KW}$ is the energy
flux in three  dimensional $\B k$-space.

Later L'vov-Nazarenko (LN)~\cite{LN} criticised the KS assumption of
locality and concluded that the leading contribution to the energy transfer
comes from  a six waves scattering in which two wave vectors (from the same
side) have wavenumbers of the order of $1/\ell$.
%process the $(3 \Leftrightarrow 3)$-wave scattering in which two wave
%vectors (from one side) are about $1/\ell$.
LN concluded that the spectrum is

 \begin{equation}\label{LN}
	 E\Sp{LN}\Sb{LN}(k)\simeq \Lambda \,   \varepsilon^{1/3}\Sb{KW} \kappa\, \ell^{-4/3}k^{-5/3},\  \  \mbox{(LN spectrum).} \\
	 \end{equation}

\noindent
This KS vs LN controversy triggered an intensive debate (see e.g.
Refs~\cite{LN-debate1,LN-debate2,LN-debate3,KS-debate,Sonin,LN-Sonin}),
which is outside the scope of this article.  We only mention that
the three--dimensional energy spectrum $E_{KW}(k)$ can be related
to the one--dimensional amplitude spectrum $A_{KW}(k)$ by
$E_{KW}(k) \sim \hbar \omega n$ where $\omega(k) \sim k^2$ is
the angular frequency of a Kelvin wave of wavenumber $k$, $\hbar \omega$ the
energy of one quantum, and $n \sim A_{KW}$ the
number of quanta; therefore, in terms of the Kelvin waves amplitude spectrum
(which is often reported in the literature and can be numerically computed), the two predictions are respectively
$A_{KW}^{KS} \sim k^{-17/5} = k^{-3.40}$ (KS) and
$A_{KW}^{LN} \sim k^{-11/3} = k^{-3.67}$ (LN).
The two predicted exponents
(-3.40 and -3.67) are very close to each other;
indeed VFM simulations~\cite{Carlo4}   could not distinguish them
(probably because the numerics were not in the sufficiently weak regime
of the theory in terms of ratio of amplitude to wavelength).
Nevertheless, more recent GPE simulations by Krstulovic~\cite{Krs} based on
long time integration of Eq.~\eqref{GPE} and averaged over
initial conditions (slightly deviating from a straight line)
support the LN spectrum~\eqref{LN}.

At finite temperature, it was shown
in Ref.~\cite{BoueLvovProcaccia:EPL2012} that the Kelvin wave
spectrum~\eqref{LN} is suppressed by mutual friction for $k>k_*$,
reaching core scale ($k_* \xi \approx 1$) at $T\simeq 0.07\,$K
and fully disappears
at $T\simeq 1\,$K, when $k_*\ell \approx 1$.

\section{\label{VIII}
  Models: near the intervortex scale.}
%\section{  7. Energy spectra
% near the intervortex scale}

The region of the spectrum near the intervortex scale
$k \ell \approx 1$ is difficult because both eddy-type motions and
Kelvin waves are important, and the discreteness of the superfluid
vorticity prevents direct application of the tools of
classical hydrodynamic. Nevertheless, some progress can be made:
Sect.~\ref{8A} presents a differential model for
the $T\to 0$ limit~\cite{LNR-2}, and  Sect.~\ref{8B} describes
a complementary truncated HVBK model~\cite{Salort-2011} designed
for the $T>1~\rm K$ temperature range.
% . Sect.~8B describes an
%alternative model which has been proposed\cite{Salort-2011},
%in which the discrete nature of superfluid vorticity
%affects the spectrum even at $k<k_{\ell}$, thus extending this
%difficult range.

\subsection{\label{8A} Differential model of $^4$He at zero temperature}
The description of superfluid turbulence for $k\ell \approx 1$  is more
complicated than $k \ell \gg 1$ because there are no well justified
theoretical approaches (like in the problem of Kelvin wave turbulence
at $k\ell \gg 1$) or even commonly accepted uncontrolled closure
approximations. Nevertheless, there some qualitative
predictions can be tested numerically and experimentally,
at least in the zero temperature limit.

Comparison \cite{LNR-1} of the
hydrodynamic spectrum~\eqref{K41} with the Kelvin wave spectrum
at $T=0$ suggests that the one dimensional
nonlinear transfer mechanisms among weakly
nonlinear Kelvin waves on individual vortex lines is less
efficient than the three--dimensional, strongly nonlinear
eddy-eddy energy transfer. The consequence is an
energy cascade stagnation at the crossover between the collective
eddy-dominated scales and the single vortex wave-dominated scales.
Ref.~\cite{LNR-2} argued that
the superfluid energy spectrum $E(k)$ at $k\ell \approx 1$ should be
a mixture of three--dimensional hydrodynamics modes and
one--dimensional Kelvin waves motions; the corresponding spectra
should be
$$
E\Sp{HD}\!(k)\equiv  g(k \ell)E(k)\,, \quad E\Sp{KW}\!(k)\equiv [1-  g(k\ell)]E(k)\ .
$$
Here
$$ g(x)\simeq \big[1+ x^2 \exp (x)/ 4\pi (1+x) \big]^{-1}%%
$$
is the``blending" function which  was found~\cite{LNR-2}
by calculating the energies of correlated and uncorrelated motions
produced by a system of $\ell$-spaced wavy vortex lines.

The total energy flux, $\varepsilon(k)$ arising from hydrodynamic and
Kelvin-wave contributions, was modelled~\cite{LNR-2} by dimensional
reasoning in the differential approximation, similar to Eq.~\eqref{Leith-n}:
for $k \to 0$ the energy flux is purely hydrodynamic
and $E(k)$ is given by Eq.~\eqref{flux1}, while for $k \to \infty$
it is purely supported by Kelvin waves and $E(k)$ is given by Eq.~\eqref{LN}.
This approach leads to the   ordinary differential equation
$\varepsilon(k)=$ constant, which was solved numerically. The
predicted energy spectra $E(k)$ for different values of $\Lambda$
are shown in Fig.~\ref{f:3}C, exhibit a bottleneck energy
accumulation $E(k)\propto k^2$ in agreement with Eq.~\eqref{flux1}.

%The spectrum of
%$| \omega_s |$ has been computed using the truncated continuous
%model and found temperature
%dependent within the Kolmogorov cascade \cite{Salort-2011}.
%In the low temperature limit, superfluid spectra are compatible
%with a $k^{-5/3}$ scaling (same as experiment) while they tend
%toward a flat spectrum at high temperature (same as the classical
%spectra) as illustrated by Figure~\ref{fig:vldTrunc}.
%Interestingly,
%the spectra of $| \omegab n |$ remains flat (and low) at all
%temperatures which suggest that the $k^{-5/3}$  scaling within the
%Kolmogorov cascade ($10 \lesssim k \lesssim 50$ on the figure
%\ref{fig:vldTrunc}) is directly related to the accumulation of
%superfluid thermalized excitations at small scales and low temperature
%(see the corresponding Figure \ref{f:3}-A), in agreement with the
%first model presented above. We note that some features of these
%simulations are also consistent with the second model,
%in particular the temperature dependence of both scalings
%and VLD spectra intensities.

\subsection{\label{8B} Truncated HVBK model of $^4$He at finite temperature}
%\magenta{The model~\cite{LNR-1,LNR-2} presented in Sec.~8A predicts
%the energy accumulation at due to small efficiency of Kelvin waves
%for the energy transfer for $k\ell>1$ at $T=0$.
Recently a model~\cite{Salort-2011} has been proposed that
accounts for the fact that  (according to numerical
evidence~\cite{Schwarz1988,Mineda:JLTP2012,Kivotides-2011} and analytical
estimates~\cite{Samuels1999,Vinen:2002,BoueLvovProcaccia:EPL2012})
small scales excitations ($R < \ell$), such as Kelvin waves and
isolated rings, are fully damped for $T \gtrsim 1\,$K.
Thus, at these temperatures, the energy flux $\varepsilon_s(k)$
should be very small at scales $k \gtrsim k_\ell$.
The idea~\cite{Salort-2011} was to use the HVBK Eqs.~\eqref{NSE}
but truncating the superfluid beyond a cutoff wavenumber
$k_{\ell^\star}=\beta k_\ell$,  where
$\beta$ is a fitting  parameter of order one. Obviously,
a limitation of this model is the abruptness of the truncation
(a more refined model could incorporate a smoother closure which
accounts for the dissipation associated with vortex reconnections and
the difference between ${\cal L}$ and $\vert \omega_s \vert / \kappa$).

Direct numerical simulations of this truncated HVBK model for temperatures
$1~\rm K < T < T_{\lambda}$ with $\beta=0.5$ confirmed the KO--41
scaling of the two locked fluids in the range $k_D \ll k <k\sb{meso}$
(see Fig.~\ref{f:3}A). At smaller scales, an intermediate (meso) regime
$k\sb{meso} < k < k_\ell$
was found that expands as the temperature is lowered \cite{Salort-2011}.
Apparently, superfluid energy, cascading from larger length
scales, accumulates beyond $k\sb{meso}$. At the lowest temperatures,
this energy appears to thermalize, approaching equipartition
with $E_s(k) \propto k^2$, as shown by the red curve of
Fig.~\ref{f:3}A. The process saturates when the friction coupling with
the normal fluid becomes strong enough to balance the incoming
energy flux $\varepsilon(k\sb{meso})$. In physical space, this mesoscale
thermalization should manifest itself as a randomisation of the vortex
tangle. The effect is found to be strongly temperature
dependent\cite{Roche:EC13}:
$ k\sb{meso} \propto  k_\ell\sqrt{ \rho\sb n/ \rho}$.

The truncated hydrodynamic model reproduces the decreasing spectrum
of the vortex line density fluctuations at small $k$ and reduces
to the classical spectrum in the $T \rightarrow T_\lambda$ limit.
This accumulation of thermalized
superfluid excitations at small scales and finite temperature was
predicted by an earlier model developed to
interpret vortex line density spectra \cite{Roche-2008}.

\section{Conclusions}
We conclude that, at large hydrodynamic scales $k_D \ll k \ll k_{\ell}$,
the evidence for the KO--41 $k^{-5/3}$ scaling of the superfluid
energy spectrum  which arises from experiments, numerical simulations and
theory (across all models used) is strong and consistent, and appears to be
independent of temperature (including the limit of zero temperature
in the absence of the normal fluid \cite{Nore,Kobayashi,
Araki,Baggaley-structures}).
This direct spectral evidence is
also fully consistent with an indirect body of evidence arising from
measurements of the kinetic energy dissipation
(\cite{Walstrom1988a,Rousset:Pipe1994,Abid1998,Fuzier:2001p235,Salort-2010})
and vortex line density decay \cite{Skrbek2001,Niemela:2005p260}
in turbulent helium flows.
 The main open issue is the existence of
vortex bundles \cite{Baggaley-structures,Sherwin-2012} predicted by the
VFM, for which there is no direct experimental observation yet.
Intermittency effects, predicted by shell models~\cite{Boue:PRL2013},
also await for experimental evidence.

What happens at mesoscales just above $k \approx k_{\ell}$
is less understood. The differential model (at $T=0$, Sect.~\ref{8A}) and
the truncated HVBK model (at finite $T$, Sec.~\ref{8B}), predict an
upturning of the spectrum (temperature-dependent for the latter model)
in this region of $k$-space.
If confirmed
by the experiments and the VFM model, this  would signify the striking
appearance of quantum effects at scales larger than $\ell$.
Further insight could arise from better understanding
of fluctuations of the vortex line density.
 It is worth
noticing that similar macroscopic manifestation of the singular nature of
the superfluid vorticity was also predicted for the pressure
spectrum \cite{Vassilicos-pressure}.

At length scales of the order of $\ell$ and less than $\ell$
the situation is even less clear.
This regime is very important at the lowest temperatures, where the Kelvin
waves are not damped, and energy is transferred from
the eddy--dominated, three--dimensional
Kolmogorov-Richardson cascade into a Kelvin wave cascade on individual
vortex lines, until the wavenumber is large enough that energy is
radiated as sound. The main open issues which call for better
understanding concern the cross--over
and
%the possibility of a bottleneck between the two cascades,
more elaborated description of the bottleneck energy
accumulation around $\ell$ in the
wide temperature range from 0 to about $T_\lambda$
and the role of vortex reconnections in the
strong regime (large Kelvin wave amplitudes compared to wavelength)
of the cascade. At the moment, there is
much debate on these problems but
no direct experimental evidence for these effects.
It is however encouraging that the most recent GPE simulations \cite{Krs}
in the weak regime (small amplitude compared to
wavelength) seem to agree with theoretical predictions.

%\blue{The other spectrum which is explored in turbulent superfluid helium
%is the vortex line density spectrum. Here the situation is less certain and
%we refer the reader to dedicated experiments \cite{Roche-2007,Lancaster}
%numerics \cite{Fujiyama,Salort-2011,Baggaley-fluctuations,Laurie-2012}
%and models \cite{Roche-2008,Nemirovskii:PRB2012}.}
%Finally, we the study of the higher order statistics
%\cite{Tabeling,Salort-EPL-2012,SalortIntermi:ETC13_2011,Boue:PRL2013}
%is still at an early stage.
% We hope that this work, which attempts to
%establish a better understanding of the superfluid analogue of
%the KO--41 regime, will launch a systematic investigation of
%higher order effects \red{and intermittency in quantum fluids.}

%{\color{red} Test colour}

\begin{acknowledgments}
C.F.B is grateful to EPSRC for financial support.
P.-E.R. acknowledges %an enlightening discussion on $^3$He quasi particles with M. Jackson,
numerous exchanges with E. L\'ev\^eque and financial support from
ANR-09-BLAN-0094 and from la R\'egion Rh\^one-Alpes.
\end{acknowledgments}

%\end{article}

\begin{thebibliography}{10}

%1
\bibitem{Annett}
A.F. Annett (2004) {\em Superconductivity, superfluids and condensates}
(Oxford University Press, Oxford).

%2
\bibitem{Skrbek-Sreeni-2012}
L. Skrbek and K.R. Sreenivasan (2012),
{\em Developed quantum turbulence and its decay}, Phys. Fluids (24): 011301.

%3
\bibitem{Hulton}
C.F. Barenghi, S. Hulton and D.C. Samuels (2002),
{\em Polarization of superfluid turbulence},
Phys. Rev. Lett. (89): 275301.

% 4
\bibitem{Laurie-2012}
A.W. Baggaley, J. Laurie, and C.F. Barenghi (2012),
{\em Vortex-density fluctuations, energy spectra, and vortical regions
in superfluid turbulence},
Phys. Rev. Lett. (109): 205304.

%5
\bibitem{Donnelly}
R.J. Donnelly (1991) {\em Quantised Vortices In Helium II}
(Cambridge University Press, Cambridge, UK).

%6
\bibitem{BDV1982}
C.F. Barenghi, R.J. Donnelly, and W.F. Vinen (1982),
{\em Friction on quantized vortices in helium II}
J. Low Temp. Phys. (52): 189--247.

%7
\bibitem{DB} R.J. Donnelly and C.F. Barenghi (1998),
{\em The observed properties of liquid helium at the saturated
vapor pressure},
J. Phys. Chem.  (6): 1217--1274.

% removed
%\bibitem{Kov47}  L. Kovasznay (1947)\red{MISS TITLE}
%J. Aeronaut.Sci. (15): 745.

%8
\bibitem{Leith67} C. Leith (167),
{\em Diffusion approximation to inertial energy transfer in isotropic
turbulence},
Phys. Fluids (10): 1409--1416.

%9
\bibitem{Nazar-Leith} C. Connaughton and S. Nazarenko (2004),
{\em Warm Cascades and Anomalous Scaling in a Diffusion Model of Turbulence},
Phys. Rev. Lett. (92): 044501.

%10
\bibitem{Frisch}
U. Frisch (1995),
{\em Turbulence. The legacy of A.N. Kolmogorov}
(Cambridge University press, Cambridge, UK).

%11
\bibitem{Baggaley-structures}
A.W. Baggaley, C.F. Barenghi, A. Shukurov, and Y.A. Sergeev (2012),
{\em Coherent vortex structures in quantum turbulence},
Europhys. Lett. (98): 26002.

%
\bibitem{Vinen2001}
W.F. Vinen (2001),
{\em Decay of superfluid turbulence at very low temperature: the radiation
of sound from a Kelvin wave on a quantized vortex},
Phys. Rev. B (64): 134520.

%11
\bibitem{Fisher-Udine}
S.N. Fisher (2008),
{\em Turbulence experiments in superfluid 3He at very low
temperatures},
in {\em Vortices and Turbulence at Very Low Temperatures},
edited by C.F. Barenghi and Y.A. Sergeev, CISM Courses and Lectures,
vol. 501, Springer Verlag (2008), 157--257.

%12
\bibitem{Sherwin-2012}
A.W. Baggaley, L.K. Sherwin, C.F. Barenghi, and Y.A. Sergeev (2012),
{\em Thermally and mechanically driven quantum turbulence in helium II},
Phys. Rev. B (86): 104501.

%13
\bibitem{Nemirovskii:PhysRep2013}
S.~K. Nemirovskii (2013),
{\em Quantum turbulence: Theoretical and numerical problems}
Physics Reports, (524): 85--120.

%14
\bibitem{Tabeling}
J. Maurer, and P. Tabeling (1998),
{\em Local investigation of superfluid turbulence},
Europhys. Lett. (43): 29--34.


%15
\bibitem{Schmoranzer-2009}
D. Schmoranzer, M. Rotter, J. Sebek, and L. Skrbek (2009),
{\em Experimental setup for probing a von Karman type flow of
normal and superfluid helium},
Experimental Fluid Mechanics 2009, Proceedings of the International
Conference, 304

%16
\bibitem{SHREK}
B. Rousset et al.,
{\em The SHREK superfluid Von Karman flow experiment},
Proc. of CEC/ICMC, Anchorage USA, June 2013

%17
\bibitem{Salort-EPL-2012}
J. Salort, B. Chabaud, L{\'e}v{\^e}que E. and P.-E. Roche (2012),
{\em Energy cascade and the four-fifths law in superfluid turbulence},
Europhys. Lett. (97):  34006.

%18
\bibitem{Roche-2007}
P.E. Roche, P. Diribarne, T. Didelot, O. Francais, L. Rousseau,
and H. Willaime (2007),
{\em Vortex density spectrum of quantum turbulence},
Europhys. Lett. (77):  66002.

%19
\bibitem{Salort-2010}
J. Salort J, et al. (2010),
{\em Turbulent velocity spectra in superfluid flows},
Phys. Fluids (22): 125102.

%20
\bibitem{Salort-2011}
J. Salort, P.-E. Roche and E. L{\'e}v{\^e}que (2011),
{\em Mesoscale Equipartition of kinetic energy in Quantum Turbulence},
Europhys. Lett. (94): 24001.

%21
\bibitem{SalortIntermi:ETC13_2011}
J.~Salort, B.~Chabaud, E.~L{\'e}v{\^e}que, and P.-E. Roche (2011).
\newblock Investigation of intermittency in superfluid turbulence.
\newblock In {\em J. Phys.: Conf. Ser.}, volume 318 of {\em Proceedings of the
  13th EUROMECH European Turbulence Conference, Sept 12-15, 2011, Warsaw}, page
  042014. IOP Publishing.

%22
\bibitem{Salort-RSI-2012}
J. Salort, A. Monfardini, and P.-E. Roche (2012),
{\em Cantilever anemometer based on a superconducting
micro-resonator: Application to superfluid turbulence},
Rev. Sci. Instrum. (83): 125002.

%%%%%%%%%%%%%%%%%%%%%%%%%%%%%%%%%%%%%%%%%%%%%%%%%%%%%%%%%%%%%%%%
% SECTION 4

%23
\bibitem{Reatto}
S.A. Vitiello, L. Reatto, G.V. Chester, and M.H. Kalos (1996),
{\em Vortex line in superfluid 4He: A variational Monte Carlo calculation},
Phys. Rev. B (54): 1205--1212.

% removed
%\bibitem{Pomeau:1993p3780}
%Y.~Pomeau and S.~Rica (1993),
%{\em Model of superflow with rotons}
%Phys. Rev. Letters (71) 247--250.

% removed
%\bibitem{Nugzar2008}
%C.F. Barenghi, Y.A. Sergeev, and N. Suramlishvili (2008),
%{\em Ballistic propagation of thermal excitations near a vortex in
%superfluid 3He-B} Phys. Rev. B (77):  104512.


%24
\bibitem{Schwarz1988}
K.W. Schwarz (1988),
{\em Three-dimensional vortex dynamics in superfluid 4He:
Homogeneous superfluid turbulence},
Phys. Rev. B (38):  2398--2417.

%25
\bibitem{Saffman}
P.G. Saffman (1992), {\em Vortex Dynamics}
(Cambridge University Press, Cambridge, UK).

%26
\bibitem{Baggaley-tree}
A.W. Baggaley and C.F. Barenghi (2012),
{\em Tree method for quantum vortex dynamics},
J. Low Temp. Phys. (66): 3--20.

%27
\bibitem{Baggaley-recon}
A.W. Baggaley (2012),
{\em The sensitivity of the vortex filament method to different
reconnections models},
J. Low Temp. Phys. (68): 18.

%28
\bibitem{Zuccher}
S. Zuccher, M. Caliari, and C.F. Barenghi (2012),
{\em Quantum vortex reconnections},
Phys. Fluids (24): 125108.

%29
\bibitem{Kivotides-ring}
D. Kivotides, C.F. Barenghi, and D.C. Samuels (2000),
{\em Triple vortex ring structure in superfluid helium II},
Science (290): 777--779.

%30
\bibitem{Kivotides-2011}
D. Kivotides (2011),
{\em Spreading of superfluid vorticity clouds
in normal-fluid turbulence},
J. Fluid Mech. (668): 58--75.

%31
\bibitem{Hall-Vinen}
H.E. Hall and W.F. Vinen (1956),
{\em The rotation of liquid helium II. II. The theory of mutual
friction in uniformly rotating helium II.}
Proc. Roy. Soc. London A (238): 215--234.

%32
\bibitem{Hills-Roberts}
R.N. Hills and P.H. Roberts (1977),
{\em Superfluid mechanics for a high density of vortex lines},
Arch. Rat. Mech. Anal. (66): 43--71.

%33
\bibitem{Barenghi-Couette}
C.F. Barenghi (1992),
{\em Vortices and the Couette flow of helium II},
Phys. Rev. B (45): 2290--2293.

%34
\bibitem{Henderson}
K.L. Henderson, C.F. Barenghi and C.A. Jones (1995),
{\em Nonlinear Taylor Couette flow of helium II},
J. Fluid Mech. (283):  329--340.

%%%%%%%%%%%%%%%%%%%%%%%%%%%%%%%%%%%%%%%%%%%%%%%%%%%%%%%%%%%%%%%%
% SECTION 5

%35
%\bibitem{Nugzar2013}
%N. Suramlishvili, C.F. Barenghi, Y.A. Sergeev, S.N. Fisher, V. Tsepelin, and G. Pickett,
%{\em Interpretation of quasi particle scattering measurements in $^3$He-B: a three--dimensional numerical analysis},
%in preparation.

%36
%\bibitem{VanSciver2009}
%S.W. Van Sciver and C.F. Barenghi (2009),
%{\em Visualization of quantum turbulence},
%in {\em Progress of Low Temperature Physics}, 247--303,
%ed. by M. Tsubota and W.P. Halperin, Elsevier.

%37
%\bibitem{Fonda}
%E. Fonda, D.P. Meichle, N.T. Ouellette, S.H., K.R. Sreenivasan, and D.P. Lathrop,
%{\em Visualization of Kelvin waves on quantum vortices}
%arXiv:1210.5194 (2012).

%38
%\bibitem{Guo}
%W. Guo, S.B. Cahn, J.A. Nikkel, W.F. Vinen, and D.N. McKinsey (2010),
%{\em Visualization Study of Counterflow in Superfluid 4He using Metastable Helium Molecules},
%Phys. Rev. Lett. (105): 045301.

% 39
%\bibitem{LaMantia:JFM2013}
%M.~La~Mantia, D.~Duda, M.~Rotter and L.~Skrbek (2013).
%{\em Lagrangian accelerations of particles in superfluid turbulence}
%J. Fluid Mech. (717): R9 doi:10.1017/jfm.2013.31.

% 40
\bibitem{Nore}
C. Nore, M. Abid, and M.E. Brachet (1997),
{\em Kolmogorov turbulence in low temperature superflows},
Phys. Rev. Lett. (78): 3896--3899.

% 41
\bibitem{Kobayashi}
M. Kobayashi, and M. Tsubota (2005),
{\em Kolmogorov spectrum of superfluid turbulence: numerical analysis
of the Gross-Pitaevskii equation with a small-scale dissipation},
Phys. Rev. Lett. (94): 065302.

% 42
\bibitem{Sasa-2011}
N. Sasa, T. Kano, M. Machida, V. S. L'vov, O. Rudenko and M. Tsubota (2011),
\em{Energy spectra of quantum turbulence: Large-scale simulation and modelling},
Phys. Rev. B (84): 054525.

% 43
\bibitem{LNR-2}V. S. L'vov, S. V. Nazarenko, O. Rudenko (2008),
{\em Gradual eddy-wave crossover in superfluid turbulence},
J. Low Temp. Phys. (153): 140--161.

% 44
\bibitem{Araki}
T. Araki, M. Tsubota, and S.K. Nemirovskii (2002),
{\em Energy Spectrum of Superfluid Turbulence with No Normal-Fluid Component},
Phys. Rev. Lett. (89): 145301.

%\bibitem{Baggaley-structures}
%A.W. Baggaley, C.F. Barenghi, A. Shukurov, and Y.A. Sergeev (2012),
%{\em Coherent vortex structures in quantum turbulence},
%Europhys. Lett. (98): 26002.

% 46
\bibitem{Tsubota-PLTP}
M. Tsubota and M. Kobayashi (2009),
{\em Energy spectra of quantum turbulence}, in
 {\em Progress of Low Temperature Physics}, 1--43,
ed. by M. Tsubota and W.P. Halperin, Elsevier.

% 47
\bibitem{Proukakis:2008}
N. P. Proukakis and B. Jackson (2008),
{\em Finite-temperature models of Bose-Einstein condensation},
J Phys B-At Mol Opt {\bf 41} (2008): 203002.

% 48
\bibitem{Brachet:CRPhys2012}
M. E. Brachet (2012),
{\em Gross-Pitaevskii description of superfluid dynamics at finite temperature: A short review of recent results},
C. R. Physique (13): 954.

% 49
\bibitem{Allen2013}
A.J. Allen, E. Zaremba, C.F. Barenghi, and N.P. Proukakis (2013),
{\em Observable vortex properties in finite-temperature Bose gases}
Phys. Rev. A (87): 013630.

% 50
\bibitem{Krstulovic:PRB2011}
G.~Krstulovic and M.~Brachet (2011),
{\em  Anomalous vortex-ring velocities induced by thermally excited Kelvin
  waves and counterflow effects in superfluids.}
Phys. Rev. B (83): 132506.

% 51
\bibitem{Aarts}
R.G.K.M. Aarts, and A.T.A.M. de Waele (1994),
{\em Numerical investigation of the flow properties of He II},
Phys. Rev. B (50) 10069--10079.

% 52
\bibitem{Kivotides-2002}
D. Kivotides, C.J. Vassilicos, D.C. Samuels, and C.F. Barenghi (2002),
{\em Velocity spectra of superfluid turbulence},
Europhys. Lett. (57): 845--851.

% 53
\bibitem{Osborne-2006}
D.R. Osborne, J.C. Vassilicos, K. Sung, and J.D. Haigh (2006),
Fundamentals of pair diffusion in kinematic simulations of turbulence,
Phys. Rev. E (74): 036309.

% 54
\bibitem{Morris}
K. Morris, J. Koplik, and D.W.I. Rouson (2008),
{\em Vortex locking in direct numerical simulations of quantum turbulence},
Phys. Rev. Lett. (101): 015301.

% 55
\bibitem{Kivotides-2006}
D. Kivotides (2006),
{\em Coherent structure formation in turbulent thermal superfluids},
Phys. Rev. Lett. (96): 175301.

% 56
\bibitem{LNR-1}V. S. L'vov, S. V. Nazarenko, O. Rudenko (2007),
{\em Bottleneck crossover between classical and quantum superfluid turbulence},
Phys. Rev. B {76}, 024520.

% 57
\bibitem{Adachi}
H. Adachi, S. Fujiyama, and M. Tsubota (2010),
{\em Steady state counterflow turbulence: simulation of vortex filaments using the full Biot-Savart law.}
Phys. Rev. B (81): 104511.


% 58
\bibitem{Roche-2008}
P.E. Roche, and C.F. Barenghi (2008),
{\em Vortex spectrum in superfluid turbulence: interpretation of a
recent experiment},
Europhys. Lett. (81): 36002.

% 59
\bibitem{Merahi:2006}
L.~Merahi, P.~Sagaut, and Z.~Abidat (2006),
{\em A closed differential model for large-scale motion in {HVBK} fluids},
Europhys. Lett. (75): 757.

% 60
\bibitem{Roche2fluidCascade:EPL2009}
P.-E. Roche, C.~F. Barenghi, and E.~L{\'e}v{\^e}que (2009),
{\em Quantum turbulence at finite temperature: the two-fluids cascade},
Europhys. Lett.  (87): 54006.

% 61
\bibitem{Tchoufag:POF2010}
J.~Tchoufag and P.~Sagaut (2010),
{\em Eddy damped quasi normal Markovian simulations of superfluid
  turbulence in helium II} ,
Phys. Fluids, (22): 125103.

% 62
\bibitem{Wacks-2011}
D. H. Wacks and C. F. Barenghi (2011),
\em{Shell model of superfluid turbulence},
Phys. Rev. B (84): 184505.

% 63
\bibitem{Boue:PRL2013}
L.~Bou\'e, V.~L'vov, A.~Pomyalov, and I.~Procaccia (2013)r,.
{\em  Enhancement of intermittency in superfluid turbulence.}
 Phys. Rev. Lett. (110): 014502.

% 64
\bibitem{Boue:PRB2012}
L.~Bou\'e, V.~L'vov, A.~Pomyalov, and I.~Procaccia (2012)r,.
{\em  Energy spectra of superfluid turbulence in $^3$He}
 Phys. Rev.B (85): 104502.

%%%%%%%%%%%%%%%%%%%%%%%%%%%%%%%%%%%%%%%%%%%%%%%%%%%%%%%%%%%%%%%%
% SECTION 6

% 65
\bibitem{LNV}
V.~S. L'vov, S.~V. Nazarenko, G.~E. Volovik (2004),
{\em Energy spectra of developed superfluid turbulence},
JETP Letters (80) 479.

% 66
\bibitem{He3spectra}
L. Bou{\'e}, V.S. L'vov, A. Pomyalov, and I. Procaccia (2012),
{\em Energy spectra of superfluid turbulence in $^3$He}
Phys. Rev. B (85):  104502.

% 67
\bibitem{LNS} V.~S. L'vov, S.~V. Nazarenko, L.~Skrbek (2006),
{\em Energy Spectra of
Developed Turbulence in Helium Superfluids},
J. Low Temp. Phys. (145): 125.

%%%%%%%%%%%%%%%%%%%%%%%%%%%%%%%%%%%%%%%%%%%%%%%%%%%%%%%%%%%%%%%%
% SECTION 7


% 68
\bibitem{Svi95}  B. V. Svistunov (1995),
{\em Superfluid turbulence in the low-temperature limit},
Phys. Rev. B (52): 3647.

% 69
\bibitem{Carlo2} D. Kivotides, J.C. Vassilicos, D.C. Samuels,
and C.F. Barenghi (2001),
{\em Kelvin Waves Cascade in Superfluid Turbulence},
Phys. Rev. Lett. (86): 3080.

%70
\bibitem{Tsubota3} W. F. Vinen, M. Tsubota, and A. Mitani (2003),
{\em Kelvin-Wave Cascade on a Vortex in Superfluid 4He at a Very Low
Temperature},
Phys. Rev. Lett. (91): 135301.

%71
\bibitem{KS}
E. Kozik, B.V.  Svistunov (2004),
Phys. Rev. Lett. (92): 035301.

%72
\bibitem{LN}
V.S. L'vov, S. Nazarenko (2010),
Spectrum of Kelvin-wave turbulence in superfluids
Pisma v ZhETFf (91): 464.

%73
\bibitem{LN-debate1}
J. Laurie, V.S. L'vov, S. Nazarenko, O. Rudenko (2010),
{\em Interaction of Kelvin waves and nonlocality of energy
transfer in superfluids},
Phys. Rev. B. (81): 104526.

%74
\bibitem{LN-debate2}
V.V. Lebedev, V.S. L'vov (2010),
{\em Symmetries and interaction coefficient of Kelvin waves}
J. Low Temp. Phys, (161): 548--554.

%75
\bibitem{LN-debate3}
V.V. Lebedev, V.S. L'vov, S.V. Nazarenko (2010),
{\em Reply: On role of symmetries in Kelvin wave turbulence}
J. Low Temp. Phys, (161): 606--610.

%76
\bibitem{KS-debate}
E. Kozik, B.V. Svistunov (2010),
{\em Geometric symmetries in superfluid vortex dynamics},
Phys. Rev. B (82):(R) 140510.

%77
\bibitem{Sonin}
E.B. Sonin (2012),
{\em Symmetry of Kelvin-wave dynamics and the Kelvin-wave cascade
in the T=0 superfluid turbulence},
 Phys. Rev. B (85): 104516.

%78
\bibitem{LN-Sonin} V.S. L'vov, S.V. Nazarenko (2012),
{\em Comment on Symmetry of Kelvin-wave dynamics and the Kelvin-wave
cascade in the T=0 superfluid turbulence},
Phys. Rev. B (86): 226501


%79
\bibitem{Carlo4}
A.W. Baggaley and C.F. Barenghi (2011),
{\em Spectrum of turbulent Kelvin-waves cascade in superfluid helium},
Phys. Rev. B (83): 134509.

%80
\bibitem{Krs}G. Krstulovic (2012),
{\em Kelvin-wave cascade and dissipation in low-temperature superfluid
vortices}
Phys. Rev. E (86): 055301.

% 81
\bibitem{BoueLvovProcaccia:EPL2012}
L.~Bou{\'e}, V.~L'vov, and I.~Procaccia (2012).
{\em Temperature suppression of Kelvin-wave turbulence in superfluids},
Europhys. Lett. (99): 46003.

%%%%%%%%%%%%%%%%%%%%%%%%%%%%%%%%%%%%%%%%%%%%%%%%%%%%%%%%%%%%%%%%%%%
% SECTION 8

% 82
\bibitem{Mineda:JLTP2012}
Y.~Mineda, M.~Tsubota, and W.~Vinen (2012).
{\em Decay of counterflow quantum turbulence in superfluid 4He},
J. Low Temp. Phys.  1 10.1007/s10909-012-0800-7

% 83
\bibitem{Samuels1999}
D.~C. Samuels and D.~Kivotides (1999),
\em{A damping length scale for superfluid turbulence}
Phys. Rev. Lett. (83): 5306.

% 84
\bibitem{Vinen:2002}
W.~F. Vinen and J.~J. Niemela (2002),
{\em  Quantum turbulence},
J. Low Temp. Phys. (128): 167.

% 85
\bibitem{Roche:EC13}
P.-E. Roche (2013),
{\em Energy spectra and characteristic scales of quantum turbulence
investigated by numerical simulations of the two-fluid model},
To appear in the Proc. of the 14th EUROMECH European Turbulence
Conference, Sept 1-4, 2013, Lyon.

% 86
\bibitem{Vassilicos-pressure}
D. Kivotides, J.C. Vassilicos, C.F. Barenghi, M.A.I. Khan,
and D.C. Samuels (2001),
\em{Quantum signature of superfluid turbulence}
Phys. Rev. Lett. (87): 275302.


%\bibitem{Fujiyama}
%S.Fujiyama and M. Tsubota (2010),
%{\em Vortex Line Density Fluctuations of Quantum Turbulence},
%J. Low Temp. Phys. (158):  428.

%\bibitem{Baggaley-fluctuations}
%A.W. Baggaley, and C.F. Barenghi,
%{\em Vortex density fluctuations in quantum turbulence},
%Phys. Rev. B 84 (2011), 020504(R).


%\bibitem{Nemirovskii:PRB2012}
%S.~K. {Nemirovskii} (2012).
%{\em Fluctuations of the vortex line density in turbulent flows of quantum
%  fluids},
%Phys. Rev. B, (86): 224505.

\bibitem{Sherwin-2013}
LK Sherwin, AW Baggaley and CF Barenghi,
{\em in preparation}

\bibitem{Walstrom1988a}
P.~Walstrom, J.~Weisend, J.~Maddocks, and S.~Van~Sciver (1988)
\em{Turbulent flow pressure drop in various {H}e {II} transfer system components.}
Cryogenics (28):101
\bibitem{Rousset:Pipe1994}
B.~{Rousset}, G.~{Claudet}, A.~{Gauthier}, P.~{Seyfert}, A.~{Martinez},
  P.~{Lebrun}, M.~{Marquet}, and R.~{van Weelderen} (1994)
\em {Pressure drop and transient heat transport in forced flow single phase helium II at high Reynolds numbers}.
Cryogenics (34):317
\bibitem{Abid1998}
M.~Abid, M.~E. Brachet, J.~Maurer, C.~Nore, and P.~Tabeling (1998)
\em{Experimental and numerical investigations of low-temperature superfluid turbulence}
Eur. J. Mech B-Fluid (17):665
\bibitem{Fuzier:2001p235}
S.~Fuzier, B.~Baudouy, and S.~W. Van~Sciver (2001)
\em{Steady-state pressure drop and heat transfer in {He II} forced flow
  at high {R}eynolds number}
Cryogenics (41):453
\bibitem{Skrbek2001}
L.~Skrbek, J.~J. Niemela, and K.~R. Sreenivasan (2001)
\em{Energy spectrum of grid-generated HeII turbulence}
Phys. Rev. E, (64):067301
\bibitem{Niemela:2005p260}
J.~Niemela, K.~Sreenivasan, and R.~Donnelly (2005)
\em{Grid generated turbulence in helium II}
J. Low Temp. Phys. (138):537


%\bibitem{Lancaster}
%D.I. Bradley,S.N. Fisher, A.M. Gu$\acute{\rm e}$nault, R.P. Haley,
%S.O'Sullivan, G.R. Pickett, and V. Tsepelin (2008),
%{\em Fluctuations and correlations of pure quantum turbulence
%in superfluid $^3$He-B},
%Phys. Rev. Lett. (101):  065302.
%
%21
%\bibitem{Nugzar2008}
%\bibitem{Ishihara2003}
%T.~Ishihara, Y.~Kaneda, M.~Yokokawa, K.~Itakura, and A.~Uno (2003),
%{\em  Spectra of energy dissipation, enstrophy and pressure by
 % high-resolution direct numerical simulations of turbulence
%in a periodic box},
%J. Phys. Soc. Japan, {72}: 983.
%


%\bibitem{Brachet:1991}
%M.~E. {Brachet} (1991),
%{\em Direct simulation of three-dimensional turbulence in the
 % Taylor-Green vortex},
%Fluid Dynamics Research (8): 1.\\
%1\\ 2 \\ 3\\ 4 \\ 5 \\ 6 \\ 7 \\ 8 \\ 9 \\ 10 \\
% 1\\ 2 \\ 3\\ 4     \\ 5     \\ 6 \\ 7 \\ 8 \\ 9 \\ 20 \\
% 1\\ 2 \\ 3\\ 4     \\ 5     \\ 6 \\ 7 \\ 8 \\ 9 \\ 30 \\
% 1\\ 2 \\ 3\\ 4     \\ 5     \\ 6 \\ 7 \\ 8 \\ 9 \\ 40 \\
%  1\\ 2 \\ 3\\ 4     \\ 5     \\ 6 \\ 7 \\ 8 \\ 9 \\ 50 \\
\end{thebibliography}
\end{document}